\documentclass[twocolumn,showpacs,preprintnumbers,amsmath,amssymb,pre]{revtex4}

\usepackage{graphicx}
\usepackage{dcolumn}
\usepackage{bm}

\begin{document}
\title{Time Resolved Correlation measurements of temporally heterogeneous dynamics}
\author{Agn\`{e}s Duri$^{1}$}
\author{Hugo Bissig$^2$}
\author{V\'{e}ronique Trappe$^2$}
\author{Luca Cipelletti$^{1}$}\email{lucacip@lcvn.univ-montp2.fr}
\affiliation{$^{1}$Laboratoire des Collo\"{\i}des, Verres et
Nanomat\'{e}riaux (UMR 5587), Universit\'{e} Montpellier 2 and
CNRS, 34095 Montpellier Cedex 5, France\\$^{2}$Department de
Physique, Universit\'{e} de Fribourg, Chemin du Mus\'{e}e 3, 1700
Fribourg, Suisse }

\date{\today}

\begin{abstract}
Time Resolved Correlation (TRC) is a recently introduced light
scattering technique that allows to detect and quantify dynamic
heterogeneities. The technique is based on the analysis of the
temporal evolution of the speckle pattern generated by the light
scattered by a sample, which is  quantified by $c_I(t,\tau)$, the
degree of correlation between speckle images recorded at time $t$
and $t+\tau$. Heterogeneous dynamics results in significant
fluctuations of $c_I(t,\tau)$ with time $t$. We describe how to
optimize TRC measurements and how to detect and avoid possible
artifacts. The statistical properties of the fluctuations of $c_I$
are analyzed by studying their variance, probability distribution
function, and time autocorrelation function. We show that these
quantities are affected by a noise contribution due to the finite
number $N$ of detected speckles. We propose and demonstrate a
method to correct for the noise contribution, based on a
$N\rightarrow \infty$ extrapolation scheme. Examples from both
homogeneous and heterogeneous dynamics are provided. Connections
with recent numerical and analytical works on heterogeneous glassy
dynamics are briefly discussed.

\end{abstract}

\pacs{82.70.-y, 42.30.Ms, 64.70.Pf, 05.40.-a}


\maketitle


\section{Introduction}
\label{SEC:Intro}

Soft glassy systems such as concentrated colloidal suspensions,
emulsions, surfactant phases, gels and foams exhibit very slow and
unusual dynamics, characterized by non-exponential relaxations of
correlation and response functions, which often depend on sample
history and may be heterogeneous both in space and time
\cite{LucaJPCM2005}. These phenomena have attracted a great
interest, in part due to their ``universal'' character. Examples
of unifying descriptions are the mode coupling theory of the
stationary average dynamics of concentrated suspensions of
particles with both repulsive or attractive interactions
\cite{DawsonPRE2001,EckertPRL2002,PhamScience2002} or, at a more
qualitative level, the concept of jamming, which rationalizes the
fluid-to-solid transition in a wide range of systems and
experimental configurations \cite{LiuNature1998}. Additionally,
soft glassy materials often exhibit intriguing similarities with
hard condensed matter glasses, such as the dependence of the
dynamics on sample history (aging phenomena
\cite{LucaPRL2000,KnaebelEPJL2000}, rejuvenation and memory
effects \cite{ViasnoffPRL2002,BonnPRL2002}) or the presence of
dynamical heterogeneity
(\cite{WeeksScience2000,KegelScience2000,MayerPRL2004}).

Light scattering is a popular means to measure the average
dynamics. In a traditional dynamic light scattering experiment,
one measures the normalized time autocorrelation function of the
scattered intensity: $\overline{g_2(\tau)} =
\overline{I_1(t)I_1(t+\tau)}/\overline{I_1(t)}^2$, where the
average $\overline{\cdot \cdot \cdot}$ is over time, $t$, and
$I_1(t)$ is the intensity collected by a single detector. The
intensity autocorrelation function provides quantitative
information on the dynamics of the sample; the way this
information is extracted depends on the experimental
configuration. In single scattering experiments,
$\overline{g_2(\tau)}$ is related to the intermediate scattering
function, $f(\tau)$, via the Siegert relation: $f(\tau) =
 \sqrt{\beta^{-1}\left ( \overline{g_2(\tau)}-1 \right )}$, where $\beta$ is the coherence factor that
depends on the size ratio between the speckle ---or coherence
area--- and the detector ~\cite{Berne1976,Goodman1975}. In the
opposite limit of strong multiple scattering, the diffusing-wave
spectroscopy (DWS) formalism ~\cite{DWSGeneral} allows the
particle mean square displacement to be calculated from
$\overline{g_2(\tau)}-1$, provided that the dynamics be spatially
and temporally homogeneous. For glassy systems the average over
time required to compute $\overline{g_2(\tau)}$ may become in
practice unfeasible, either because the dynamics is so slow that
prohibitively long experiments would be required to accumulate a
satisfactory statistics, or because the dynamics may be
non-stationary, e.g. for aging systems. Various schemes have been
introduced to address this issue, most of them based on the idea
of measuring $g_2(\tau)$ for many independent speckles and
averaging the intensity correlation function not only over time,
but also over distinct speckles. This can be done either
sequentially --e.g. by slowly rotating the sample so as to
illuminate a single detector with different speckles (interleave
\cite{MullerProgColloidPolymSci1994} or echo \cite{PhamRSI2004}
methods)-- or in parallel, e.g. by using the pixels of a
charge-coupled device camera (CCD) as independent detectors
(multispeckle method
\cite{WongRSI1993,KirschJChemPhys1996,LucaRSI1999}).

These techniques drastically reduce the required time average and
thus extend the applicability of light scattering to glassy
systems. Similarly to traditional light scattering measurements,
however, they provide information only on the average dynamics,
not on its fluctuations. However, recent theoretical and
simulation works suggest that dynamic heterogeneity is a key
feature of the slow dynamics in glassy systems. In view of the
very limited number of experiments that directly test this
behavior on soft glasses \cite{WeeksScience2000,KegelScience2000},
new experimental tools that access fluctuations of the dynamics
are needed. We have recently introduced the Time Resolved
Correlation (TRC) scheme \cite{LucaJPCM2003}, a method that allows
temporally heterogeneous dynamics to be investigated by scattering
techniques. The idea at the heart of TRC is that the temporal
evolution of the speckle pattern generated by the scattered light
will be very different for homogeneous $vs.$ heterogeneous
dynamics. For homogeneous dynamics, we expect the speckle images
to change smoothly in time. By contrast, for heterogeneous
dynamics the speckle pattern is expected to evolve
discontinuously, because the rate of change of the sample
configuration will not be constant but rather fluctuate with time.
Experimentally, the speckle images are recorded by a CCD camera
and their evolution is quantified by introducing $c_I(t,\tau)$,
the degree of correlation between images taken at time $t$ and
$t+\tau$ (a rigorous definition will be given in
sec.~\ref{SEC:measure}). Inspection of the $t$-dependence of $c_I$
at fixed $\tau$ allows temporally heterogeneous dynamics to be
discriminated from homogeneous dynamics. Indeed, in the former
case a large drop (increase) of $c_I$ is observed whenever the
dynamics is faster (slower) than average, while in the latter the
degree of correlation is constant. This method is quite general,
since it can be applied to any experimental configuration where a
multi-element detector can be used to record the speckle pattern
generated by a sample illuminated by coherent radiation. Examples
are CCD-based light scattering experiments in the single
scattering regime, both at wide \cite{KirschJChemPhys1996} and
small angle \cite{LucaRSI1999}, DWS in the transmission
\cite{ViasnoffRSI2002} or backscattering
\cite{CardinauxEurophysLett2002} geometry, and X-photon
Correlation Spectroscopy (XPCS) at small angles
\cite{DiatCOCIS1998}. In principle, the technique could also be
extended to non-electromagnetic radiation, e.g. to acoustic DWS
\cite{CowanPRE2002}.

Experiments on diluted suspensions of colloidal Brownian particles
have shown that the degree of correlation exhibits some
fluctuations even in the absence of dynamic heterogeneity
\cite{LucaJPCM2003}. As it will be shown in detail, these
fluctuations are due to statistical noise stemming from the finite
number of speckles in the CCD images. In order to exploit
quantitatively TRC data it is thus necessary to separate the
contribution to the fluctuations of $c_I$ due to the noise from
that due to dynamic heterogeneity. Although in most cases it is
not possible to directly correct the TRC time series for the
noise, we will show that it is possible to correct statistical
quantities derived from the TRC data and used to quantify the
fluctuations of $c_I$. We focus in particular on three statistical
objects: the time variance of $c_I(t,\tau)$, the probability
distribution function (PDF) of $c_I(t,\tau)$ for a fixed $\tau$,
and the time autocorrelation of the $c_I(t,\tau)$ trace itself.

The variance of $c_I$, $\sigma^2_{c_I}(\tau)$, is the lowest
moment of the data that provides information on the fluctuations.
It corresponds to the so-called dynamical susceptibility $\chi_4$
studied in many simulation and theoretical works on glassy systems
\cite{LacevicJChemPhys2003,WhitelamPRE2005,PitardPRE2005,DelGadocondmat2005}.
In a typical simulation, $\chi_4$ is the variance of the
intermediate scattering function, or a similar correlation
function describing the system's change in configuration, which is
obtained from several independent runs. Similarly,
$\sigma^2_{c_I}$ quantifies the fluctuations of the intensity
correlation function as the system evolves through statistically
independent configurations. Importantly, $\sigma^2_{c_I}$ allows
one to relate temporal dynamic heterogeneity to spatial
correlations of the dynamics. In fact, it can be shown that
$\chi_4$ is the volume integral of the spatial correlation of the
local dynamics \cite{LacevicPRE2002}. Therefore, large values of
$\sigma^2_{c_I}$ will be indicative of long-range correlations of
the dynamics. Intuitively, one can expect the variance of the
fluctuations of the dynamics to scale as the inverse number of
``dynamically independent'' regions in the scattering volume, and
thus to increase as the spatial range of the correlation of the
dynamics increases. Recent TRC experiments on a shaving cream foam
support this simple picture \cite{MayerPRL2004}.

The PDF of the TRC signal is the most complete statistical
characterization of the dispersion of the data. Any deviation from
a Gaussian shape immediately hints to heterogeneous dynamics, as
suggested by experiments on a variety of systems, including
colloidal gels
\cite{LucaJPCM2003,BissigPhysChemComm2003,SarciaPRE2005},
concentrated colloidal suspensions \cite{BallestaAIP2004} and
surfactant phases \cite{DuriFNL2005}, foams \cite{BissigPhD2004},
and granular materials \cite{CaballeroCondmat2004}. Remarkably,
the shape of the PDF of $c_I$ is often strongly reminiscent of
that obtained for similar quantities in theoretical and simulation
work on other glassy systems. An example is provided by
simulations of spin glasses, where the fluctuations of the
correlation function of the spin orientation are distributed
according to a generalized Gumbel PDF \cite{ChamonJChemPhys2004},
a probability distribution characterized by an exponential tail
strikingly similar to those reported for $c_I$ in refs.
\cite{LucaJPCM2003,BissigPhysChemComm2003,SarciaPRE2005,BallestaAIP2004,DuriFNL2005,CaballeroCondmat2004}
or shown in this paper (see figs. \ref{Fig:PDFfoam} and
\ref{Fig:cIonions}). Similar distributions are also obtained in a
variety of numerical and analytical investigations of systems with
heterogeneous dynamics, both at equilibrium and out-of-equilibrium
\cite{CluselPRE2004,MerollePNAS2005,CrisantiEurophysLett2004,PitardPRE2005}.
Indeed, it has been proposed that the Gumbel distribution arises
as a universal PDF for various quantities measured in systems with
extended spatial and/or temporal correlations
\cite{BramwellPRL2000}. Clearly, in order to compare in detail and
quantitatively the PDF measured in TRC experiments to those
obtained analytically or by simulations it is necessary to correct
the experimental data for the contribution of the measurement
noise.

The variance and the PDF of $c_I$ describe the dispersion of the
data, but are insensitive to the way the fluctuations are
distributed in time. By contrast, the time autocorrelation of the
TRC signal, which was introduced in ref. \cite{DuriFNL2005} and
which we shall term ``second correlation'', provides information
on the temporal organization of the fluctuations of the dynamics,
and sheds light on the rate and life time of rearrangement events.
The second correlation introduced here is similar to the 4-th
order intensity correlation function proposed by Lemieux and
Durian \cite{LemieuxJOSAA1999,LemieuxAPPOPT2001} and to the
multi-time correlation functions measured in nuclear magnetic
resonance experiments probing dynamical heterogeneity near the
glass transition \cite{SchmidtrohrPRL1991}. Moreover, we note that
the second correlation is the analogous, in the time domain, of
the so-called second spectrum originally introduced to investigate
non-gaussian fluctuations in the dynamics of spin glasses
\cite{WeissmanRevModPhys1993}.

In this paper, we first describe how to optimize a TRC measurement
by correcting the CCD data for the dark noise, due to the
electronic noise of the CCD, and for the uneven illumination of
the detector (sec.~\ref{SEC:measure}). We then turn to the
temporal fluctuations of $c_I$ and analyze separately the
contribution of the measurement noise, due to the finite number of
pixels, (sec.~\ref{SEC:fluctnoise}) and that due to heterogeneous
dynamics (sec.~\ref{sec:noisedeconvolution}). We illustrate our
analysis by showing TRC data obtained mainly from a dilute
suspension of Brownian particles and a shaving cream foam, two
model systems that exhibit homogeneous and heterogeneous dynamics,
respectively. We then introduce a method for correcting the
experimentally measured variance and PDF of $c_I$ for the noise
contribution. The method consist of an extrapolation scheme based
on the observation that the measurement noise vanishes in the
limit $N \rightarrow \infty$ ($N$ being the number of CCD pixels
used to calculate $c_I$), whereas the fluctuations due to
dynamical heterogeneities are independent of $N$.
In sec.~\ref{SEC:seccorr} we introduce the second correlation
and provide a working formula for correcting it for the noise
contribution. Possible artifacts that may affect the fluctuations
of $c_I$ and lead to spurious contributions to its PDF and to the
second correlation are discussed in sec.~\ref{SEC:artifacts},
together with methods to detect or correct them. In the concluding
section, we briefly discuss the connections between TRC and other
techniques that measure dynamical heterogeneities in soft glassy
systems, focussing on the advantages and the limitations of the
various methods.

\section{Time Resolved Correlation: measuring the time-dependent degree of correlation $c_I(t,\tau)$}
\label{SEC:measure}

In a TRC experiment~\cite{LucaJPCM2003} a CCD camera is used to
record, at constant rate, the speckle pattern of the light
scattered by the sample~\cite{Goodman1975}. The CCD images are
stored on the hard disk of a personal computer (PC) for later
processing. The degree of correlation, $c_I$, between speckle
patterns at times $t$ and $t+\tau$ is then calculated according to
\begin{equation}
   c_I(t,\tau) = \frac{G_2(t,\tau)}{I(t)I(t+\tau)}-1~,
   \label{Eq:cI}
\end{equation}
where $G_2(t,\tau) = \langle I_p(t)I_p(t+\tau)\rangle_p$ and $I(t)
= \langle I_p(t)\rangle_p $, with $I_p(t)$ the intensity measured
at time $t$ by the $p$-th CCD pixel and $\langle \cdot \cdot \cdot
\rangle_p $ an average over $N$ pixels. Note that the
normalization factor in Eq.~(\ref{Eq:cI}), $I(t)I(t+\tau)$, allows
to cancel out exactly any fluctuations due to changes in the laser
power. For stationary dynamics, the intensity correlation function
usually measured in a light scattering experiment,
$\overline{g_2(\tau)}$, may be obtained from $\overline{g_2(\tau)}
-1 = \overline{c_I(t,\tau)}$.

In Eq.~(\ref{Eq:cI}) the intensity value at each pixel is
required. In practice, however, the images recorded by the CCD are
affected by a pixel- and time-dependent dark (or electronic)
noise, $D_p(t)$, and, possibly, by the non-uniform illumination of
the detector, depending on the setup optics. To account for these
effects, we write the raw signal for pixel $p$, $S_p$, as $S_p(t)
= I_p(t)b_p/\langle b_p \rangle_p + D_p(t)$, where $b_p$ is a
time-independent, spatially slowly-varying function that accounts
for non-uniform illumination and $\langle b_p \rangle_p$ is
introduced so that the factor multiplying $I_p(t)$ has unit
average. Prior to each measurement, we collect 100 dark images by
covering the CCD detector with a black cap. These dark images are
used to calculate the time-averaged dark noise $\overline{D_p}$.
The raw signal is corrected by subtracting, pixel-by-pixel, the
average dark noise. To estimate $b_p$, we average over time the
dark-noise-corrected CCD signal : $b_p = \overline{S_p(t)} -
\overline{D_p}$. For experiments whose duration, $T_{\rm exp}$, is
much longer than the relaxation time of $\overline{g_2(\tau)}$,
$\tau_0$, this procedure averages out the spatial fluctuations
associated with the speckles and leads to a smooth function $b_p$,
since for each pixel the intensity fully fluctuates many times and
its time average is a good estimator of the mean intensity at a
given location of the CCD detector. By contrast, when $T_{\rm exp}
\lesssim \tau_0$, $b_p$ keeps some memory of the speckled
appearance of the instantaneous intensity distribution. In this
case, we further smooth $b_p$ by averaging it over a few adjacent
pixels. The desired intensity to be used in Eq.~(\ref{Eq:cI}) is
obtained from
\begin{equation}
   I_p(t) = [S_p(t) - \overline{D_p}]\langle b_p \rangle_p/b_p.
   \label{Eq:Icorrected}
\end{equation}
Note that $I_p(t)$ is still affected by an instanteneous noise
$\epsilon_p(t) = D_p(t) - \overline{D_p}$, since only the average
dark noise could be subtracted. By definition,
$\overline{\epsilon_p} = 0$, while the standard deviation
$\sigma_\epsilon$ of $\epsilon$ is typically of the order of 1/100
of the saturation level for an 8-bit CCD camera. When all the
relevant time scales of the dynamics are much longer than the
inverse CCD rate, $\epsilon_p(t)$ may be considerably reduced by
averaging $I_p(t)$ over a few CCD images before applying
Eq.~(\ref{Eq:cI}). In the following, we will disregard
$\epsilon_p(t)$, unless when explicitly stated.

\begin{figure}
\includegraphics{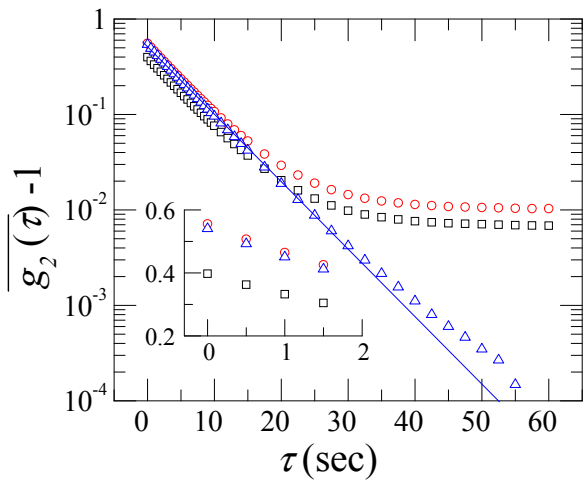}
 \caption{Semi-logarithmic plot of the intensity time autocorrelation function
 measured for a suspension of
 monodisperse Brownian particles in the single scattering regime ($\theta = 45$ deg). Data are averaged over
 $N$ = 151800 pixels and over $T_{\rm exp} = 10800$ sec. Squares: intensity autocorrelation function calculated from the CCD
 raw signal, $S_p(t)$; circles: the same data after correcting for the dark noise only;
 triangles: after correcting for both the dark noise and the uneven detector illumination. The line
 is a fit to the corrected data by a single exponential decay, the functional form of $\overline{g_2}-1$
 predicted for monodisperse Brownian particles, yielding a characteristic time $\tau_0 = 5.8$ sec.
 Inset: zoom of the small $\tau$ behavior of the intensity autocorrelation function.}
 \label{Fig:cIdarkcorrection}
\end{figure}

We show in fig.~\ref{Fig:cIdarkcorrection} how the different
corrections discussed above affect the intensity autocorrelation
function measured for a dilute suspension of monodisperse Brownian
particles. Data are collected in the single scattering geometry,
at a scattering angle $\theta = 45$ deg. The particles are
polystyrene spheres of radius $a = 265$ nm, suspended at a volume
fraction $\phi = 3.7 \times 10^{-5}$ in almost pure glycerine
cooled at 15 $^{\rm o}\rm{C}$ in order to make the dynamics slow
enough to match the limited acquisition rate of the CCD, which is
2 Hz in this experiment. Both the dark noise and the non-uniform
illumination result in a spurious increase of the base line of the
autocorrelation function (squares and circles in
fig.~\ref{Fig:cIdarkcorrection}) and lead to a change of its $\tau
\rightarrow 0$ intercept (see inset). However, the relaxation time
obtained from the small $\tau$ behavior of the intensity
autocorrelation function is essentially unaffected by the dark
noise and the non-uniform illumination. When the CCD signal is
corrected according to Eq.~(\ref{Eq:Icorrected}) (up triangles), a
single exponential decay is observed, as predicted for
monodisperse Brownian particles. For the corrected data, the base
line is limited only by the dark noise $\epsilon_p(t)$: its value
is of the order of $2 \times 10^{-4}$ and is comparable to that
obtained in traditional light scattering setups, which use a
photomultiplier tube or an avalanche photodiode as a detector.

\section{Temporal fluctuations of $c_I(t,\tau)$: the noise contribution}
\label{SEC:fluctnoise}

When disregarding the fluctuations of $\epsilon_p(t)$, the
temporal fluctuations of the degree of correlation $c_I(t,\tau)$
at a fixed lag $\tau$ have only two independent sources: the
statistical noise due to the finite number of speckles probed in
the experiment and the intrinsic fluctuations of the sample
dynamics. The first contribution is always present: we shall refer
to it as to the ``measurement noise'', not to be confused with the
dark noise discussed in Sec.~\ref{SEC:measure}. The second
contribution, on the contrary, is present only if the dynamics is
temporally heterogeneous and thus represents the physically
valuable information that we aim to extract from the fluctuations
of $c_I$. To highlight the two different contributions, we rewrite
Eq.~(\ref{Eq:cI}) as
\begin{equation}
   c_I(t,\tau) = g_2(a_1(t),...,a_k(t);\tau)-1 + n(t,\tau),
   \label{Eq:cInoise}
\end{equation}
where $n(t,\tau)$ is the measurement noise, with
$\overline{n(t,\tau)} = 0$, and $g_2(a_1(t),...,a_k(t);\tau)-1$ is
the pixel-averaged two-time intensity correlation function that
would be measured in the absence of any noise, i.e. if $c_I$ was
averaged over an infinite number of speckles. $a_1(t),...,a_k(t)$
are parameters that fluctuate with time if the dynamics are
heterogeneous, but are constant for homogeneous dynamics.

To illustrate this point, let us consider as an example of
homogeneous dynamics the single scattering measurement of the
dynamics of monodisperse Brownian particles at a scattering vector
$q$. If the temperature is carefully controlled, the diffusion
coefficient $D$ of the particles does not evolve with time and
$g_2(a_1(t),...,a_k(t);\tau)-1 = a_1\exp(-a_2\tau)$, with $a_1 =
\beta$ and $a_2 = 2q^2D$ constant \cite{Berne1976}. In this case,
the fluctuations of $c_I$ are due only to the noise $n(t,\tau)$.
By contrast, the spatially and temporally localized bubble
rearrangements in a shaving cream foam provide a simple example of
heterogeneous dynamics. For a foam, the average intensity
correlation function measured in a DWS experiment in the
transmission geometry has a shape very close to that for Brownian
particles in single scattering: $\overline{g_2}(\tau)-1 = \beta
\exp[-(2\overline{\gamma}\tau)^ {\overline{\mu}}]$, with
$\overline{\mu} \approx 0.9$ and $\overline{\gamma} =
\overline{R}r^3L^2/l^{*2}$~\cite{DurianScience1991} \footnote{For
a foam, $\overline{g_2}$ measured in the transmission geometry is
usually fitted by a more complicated expression, which can be
found in ref. \cite{DurianScience1991}. Within the experimental
uncertainty, the function of ref. \cite{DurianScience1991} can be
very well approximated by a slightly stretched exponential, which
we adopt for the sake of simplicity.}. Here, $\overline{R}$ is the
average bubble rearrangement rate per unit time and unit volume,
$r^3$ is the typical volume that is rearranged by a single event,
$L$ is the sample thickness, and $l^*$ is the photon transport
mean free path. In contrast with the Brownian suspension, however,
the degree of correlation measured for a foam fluctuates not only
because of the noise, but also because the instantaneous
rearrangement rate continuously changes due to the intermittent
nature of the bubble dynamics~\cite{BissigPhD2004,MayerPRL2004}.
Hence, for the foam $g_2(a_1(t),...,a_k(t);\tau)-1 =
a_1\exp[-(a_2(t)\tau)^{a_3(t)}]$, with $a_1 = \beta$ constant
while $a_2(t) = R(t)r^3L^2/l^{*2}$ and $a_3(t) = \mu(t)$ fluctuate
with time.

In view of the correction scheme described in secs.
\ref{sec:noisedeconvolution} and \ref{SEC:seccorr}, it is useful
to first analyze the contribution to the fluctuations of $c_I$ due
to the noise. We assume that the sample dynamics be temporally
homogeneous and stationary, as for the dilute suspension of
Brownian particles discussed above. In this case, the parameters
$a_1,...,a_k$ in Eq.~(\ref{Eq:cInoise}) are constant and the
fluctuations of $c_I$ are due only to the noise term, $n(t,\tau)$.
Since $c_I(t,\tau)$ is averaged over a large number of pixels
(typically $N \gtrsim 10000$), its temporal fluctuations at a
fixed $\tau$ are expected to be Gaussian distributed, because of
the central limit theorem (see Appendix B). Accordingly, only
$\overline{c_I}$ and the variance $\sigma^{2}_{n}(\tau)$ of the
noise $n(t,\tau)$ are needed to obtain the full probability
distribution of $c_I$: $P_{c_I}(c_I) =
(2\pi\sigma^{2}_{n})^{-1/2}\exp[-(c_I-\overline{c_I})^2/2\sigma^{2}_{n}]$.
For homogeneous dynamics, $\sigma^{2}_{n}(\tau) =
\sigma^{2}_{c_I}(\tau) \equiv
\overline{c_I(t,\tau)^2}-\overline{c_I(t,\tau)}^2$. To calculate
$\sigma^{2}_{c_I}$, we recall that the variance of a quantity $y$
that depends on the $m$ variables $x_1,...,x_m$ is given by
\begin{eqnarray}
   \sigma^{2}_{y} = \sum_{i=1}^{m} \left [\frac{\partial y}{\partial x_i}\right]_{x_i = \overline{x_i}}^2\sigma_{x_i}^2 +
   \sum_{i \neq j}  \left [\frac{\partial y}{\partial x_i}\frac{\partial y}{\partial
   x_j}\right]_{x_i = \overline{x_i} \atop x_j = \overline{x_j}} \sigma_{x_i,x_j},
   \label{Eq:var}
\end{eqnarray}
where $\sigma^2_{x_i} \equiv \overline{x_i^2} -
\overline{x_i}^{\,2}$ is the variance of $x_i$ and
$\sigma_{x_i,x_j} \equiv \overline{x_i x_j} -
\overline{x_i}\,\overline{x_j}$ is the covariance between $x_i$
and $x_j$ ($i\neq j$). The first sum accounts for the sensitivity
of $y$ to the fluctuations of the independent variables
$x_1,...,x_m$, while the second sum accounts for any correlations
between the $x_i$'s. If distinct $x_i$'s are uncorrelated,
$\overline{x_i x_j} = \overline{x_i}\,\overline{x_j}$ for $i\neq
j$ and the second sum vanishes.

By applying Eq.~(\ref{Eq:var}) to the definition  of $c_I$,
Eq.~(\ref{Eq:cI}), we find
\begin{eqnarray}
   \sigma^{2}_{c_I}(\tau)  =  1/\overline{I}^{4}\sigma^{2}_{G_2}(\tau)
   + 2\overline{G_2(\tau)}^{2}/\overline{I}^{6}\sigma^{2}_{I}
   \nonumber\\
    {} +  2\overline{G_2(\tau)}^{2}/\overline{I}^{6}\sigma_{I,J}(\tau)
    - 4\overline{G_2(\tau)}/\overline{I}^{5}\sigma_{G_2,I}(\tau),
   \label{Eq:varcI1}
\end{eqnarray}
where we have introduced the notation $J(t) \equiv I(t+\tau)$. In
writing Eq.~(\ref{Eq:varcI1}), we have used $\overline{I} =
\overline{J}$ and $\sigma_{G_2,I} =\sigma_{G_2,J}$, because the
scattered light was assumed to be stationary.

The physical origin of the fluctuations of $I$ and $G_2$,
quantified by $\sigma^{2}_{I}$ and $\sigma^{2}_{G_2}$, as well as
that of the correlation between $I$ and $J$ and between $I$ and
$G_2$, quantified by $\sigma_{I,J}$ and $\sigma_{G_2,I}$, is the
finite number of pixels over which the instantaneous intensity and
the intensity correlation are averaged. To illustrate this point,
let us consider as an example $I(t)$. As the sample evolves
through different configurations, the speckle pattern fluctuates
with a characteristic time $\tau_0$. Since the instantaneous
pixel-averaged intensity is calculated for a finite set of
speckles, different speckle patterns yield slightly different
values of $I(t)$. The larger the number of the sampled speckles,
the closer will $I(t)$ be to the ``true'' value of the average
scattered intensity, whose estimator is $\overline{I}$, and thus
the smaller will be $\sigma^{2}_{I}$. Indeed, we show in Appendix
A that $\sigma^{2}_{I} \sim N^{-1}$, as expected from the central
limit theorem . Moreover, one expects the instantaneous
pixel-averaged intensity at time $t$ to be correlated to the same
quantity measured at time $t+ \tau$, at least for $\tau \lesssim
\tau_0$, because it takes a few $\tau_0$ for the speckle pattern
to be completely renewed. Therefore, the covariance term
$\sigma_{I,J}(\tau)$ will vanish only for $\tau \gg \tau_0$. More
precisely, we expect $\sigma_{I,J}(\tau)$ to be proportional to
$\overline{G_2(\tau)}$. In fact, $\sigma_{I,J}(\tau)$ is precisely
the un-normalized correlation function measured by using the whole
CCD chip as a single detector, similarly to the case of a
traditional light scattering experiment where the detector
collects a large number of speckles.

\begin{figure}
\includegraphics{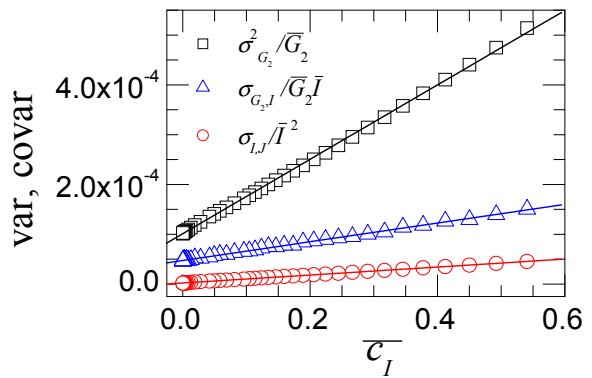}
 \caption{Parametric plot of the normalized variance and covariance terms in the r.h.s.
 of Eq.~(\ref{Eq:varcI1}), as a function of
 $\overline{c_I(\tau)}$, for the same experiment as in fig.~\ref{Fig:cIdarkcorrection}.
 The lines are linear fits to the data, suggesting the $c_I$
 dependence of Eq.~(\ref{Eq:sigmagen}).}
 \label{Fig:sigma}
\end{figure}

Similar arguments may be invoked for $\sigma^{2}_{G_2}(\tau)$ and
$\sigma_{G_2,I}(\tau)$, suggesting that the variance and
covariance factors in Eq.~(\ref{Eq:varcI1}) scale with $N^{-1}$
and depend linearly on the average correlation function:
\begin{eqnarray}
   \sigma_{x_i,x_j}(\tau) = N^{-1}[A_{x_i,x_j} + B_{x_i,x_j}\overline{c_I(\tau)}],
   \label{Eq:sigmagen}
\end{eqnarray}
where $x_i$ and $x_j$ stand for any of $I$, $J$, and $G_2$, while
$A_{x_i,x_j}$ and $B_{x_i,x_j}$ are constants whose values are
given in Appendix A.
To test the linear dependence of the variance and covariance
factors on the time-averaged correlation function, we plot
parametrically $\sigma^{2}_{G_2}(\tau)$, $\sigma_{I,J}(\tau)$, and
$\sigma_{G_2,I}(\tau)$ as a function of $\overline{c_I}(\tau)$,
for data obtained in the single scattering experiment on Brownian
particles, as shown in fig.~\ref{Fig:sigma}. In all cases the data
are very well fitted by straight lines, thus confirming the
validity of Eq.~(\ref{Eq:sigmagen}).

\begin{figure}
\includegraphics{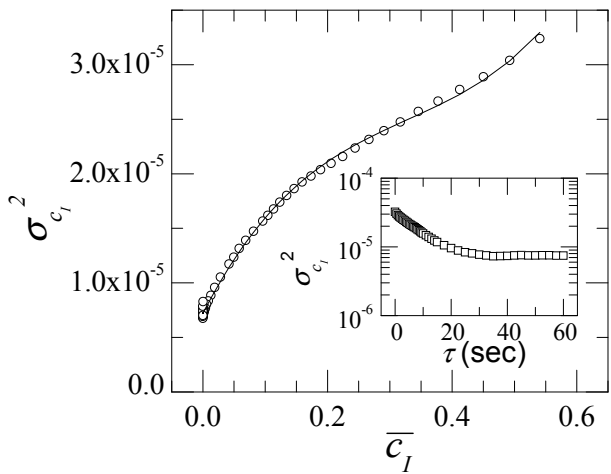}
 \caption{Main plot: parametric representation of the variance of $c_I$
 as a function of $\overline{c_I(\tau)}$ for the same experiment
 as in figs.~\ref{Fig:cIdarkcorrection} and~\ref{Fig:sigma}
 (single scattering from a dilute suspension of Brownian
 particles). The line is a fit of Eq.~(\ref{Eq:varpoly}) to the data. Inset: same data plotted $vs.$ $\tau$.}
 \label{Fig:varcI}
\end{figure}

The $\tau$ dependence of the variance of the measurement noise can
be obtained by substituting Eq.~(\ref{Eq:sigmagen}) into
Eq.~(\ref{Eq:varcI1}). Using $\overline{G_2(\tau)} =
\overline{I}^2 (\overline{c_I(\tau)}+1)$ (see Appendix B), one
finds a third order polynomial dependence of the variance of
$n(t,\tau)$ on $\overline{c_I(\tau)}$:
\begin{eqnarray}
   \sigma^{2}_{n}(\tau) = \sigma^{2}_{c_I}(\tau) = N^{-1}\sum_{l=0}^{3}\alpha_l\overline{c_I(\tau)}^{\,l},
   \label{Eq:varpoly}
\end{eqnarray}
where the coefficients $\alpha_l$ can be obtained from
$\overline{I}$, $A_{x_i,x_j}$, and $B_{x_i,x_j}$. Note that this
third-order polynomial dependence is due to the choice of the
normalization of $c_I$ (see Eq.~(\ref{Eq:cI})). If the denominator
was chosen to be $\overline{I}^2$, as in traditional light
scattering experiments, only the first term in the r.h.s. of
Eq.~(\ref{Eq:varcI1}) would be non-zero and
$\sigma^{2}_{c_I}(\tau)\propto \sigma^{2}_{G_2}$, i.e. the
fluctuations would increase linearly with $\overline{c_I(\tau)}$.
Although the normalization we have chosen leads to a more
complicated expression for $\sigma^{2}_{n}(\tau)$, we remind that
it suppresses spurious variations of $c_I$ due to fluctuations of
the incoming beam power and thus should be used in TRC
experiments. The inset of fig.~\ref{Fig:varcI} shows a
semi-logarithmic plot of $\sigma_{c_I}(\tau)$ $vs.$ $\tau$ for the
Brownian particles, confirming that the noise of $c_I$ decreases
with $\tau$, as indicated by Eq.~(\ref{Eq:varpoly}). In the main
plot, the same data are plotted parametrically as a function of
$\overline{c_I}(\tau)$. A very good agreement is found between the
experimental data and the polynomial form suggested by
Eq.~(\ref{Eq:varpoly}), as shown by the line.

\begin{figure}
\includegraphics{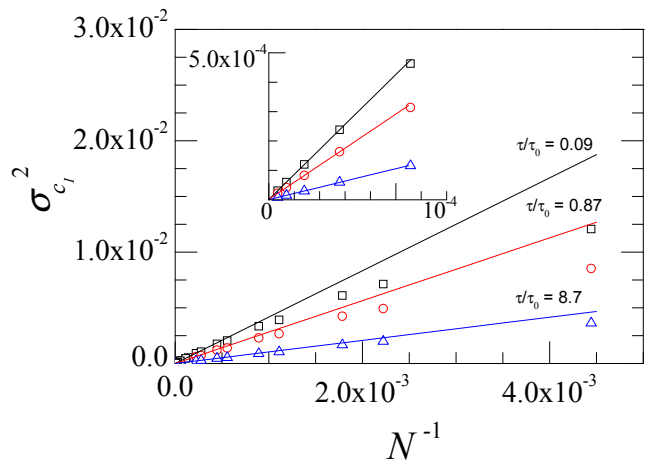}
 \caption{For the same experiment on Brownian particles as in the preceding figures: variance of $c_I$
 as a function of the inverse number
 of pixels over which $c_I$ is averaged. The lines are linear fits
 to the data for $N^{-1} \leq 1.2 \times 10^{-4}$, demonstrating that for large
 $N$ $\sigma^{2}_{c_I} \propto N^{-1}$. The inset zooms in the region near to the origin.}
 \label{Fig:varcIvsN}
\end{figure}

The $N^{-1}$ dependence of $\sigma_{n}^{2}(\tau)$ is the key
feature that will be exploited in the correction for the
measurement noise. To test this scaling, we analyze the time
series of speckle images recorded for the Brownian particle
suspension, for which $\sigma^{2}_{c_I}=\sigma^{2}_{n}$, by
processing different number of pixels. First, all pixels of each
image are processed and $c_I(t,\tau)$ and its variance
$\sigma^{2}_{c_I}(\tau)$ are calculated. Each image is then
divided into two regions of interest (ROI) of equal size. For each
ROI, $c_I$ and its variance are calculated and the values of
$\sigma^{2}_{c_I}(\tau)$ obtained for the two ROIs are averaged,
yielding the variance of $c_I$ when only $N/2$ pixels are
processed. This scheme is iterated until the size of each ROI is
reduced to 225 pixels.
Figure~\ref{Fig:varcIvsN} shows $\sigma^{2}_{c_I}(\tau)$ as a
function of the inverse number of processed pixels, $N^{-1}$, for
three time delays corresponding to 0.09, 0.87, and 8.7 times the
relaxation time of $\overline{g_2}-1$, respectively. In all cases,
the data for $N^{-1} < 1.2 \times 10^{-4}$ are very well fitted by
a straight line that goes through the origin, as shown in the
inset~\footnote{In the linear fit of $\sigma^2_{c_I}$ $vs.$
$N^{-1}$, we weight the data by the inverse of their uncertainty,
which we take to be equal to $\sigma_{c_I}$.}. This confirms that
for temporally homogeneous dynamics $\sigma^{2}_{c_I} \propto
N^{-1}$, as indicated by Eq.~(\ref{Eq:varpoly}). Note that a
deviation from this linear trend is observed at the largest
$N^{-1}$, due to edge effects. In fact, the contribution of each
pixel to $c_I$ is not completely independent from that of nearby
pixels, because the intensity of the speckle pattern is spatially
correlated over a distance of a few pixels. Pixels far from the
edges of a ROI have more nearby pixels than those on the edges;
accordingly, the statistically independent contribution to $c_I$
carried by a bulk pixel is less than that of an edge pixel. When
reducing the size of the processed ROI, the weight of edge pixels
relative to bulk pixels increases and corrections to the $N^{-1}$
scaling become increasingly apparent. These corrections are
negligible for $N \gtrsim 8000$, as seen in
fig.~\ref{Fig:varcIvsN}. We find that a similar $N^{-1}$ scaling
is obtained at all time delays (data not shown). Indeed, the
slopes of the straight line fits to $\sigma^{2}_{c_I}$ $vs.$
$N^{-1}$ provide an estimate of the proportionality coefficient
$\sum_{l=0}^{3}\alpha_l\overline{c_I(\tau)}^l$ that agrees within
$1\%$ with the value directly obtained 
when calculating $\sigma^{2}_{c_I}$ by processing the maximum
number of available pixels.

For temporally homogeneous dynamics, the statistics of the
variables $G_2(t,\tau)$ and $I(t)$ in Eq.~(\ref{Eq:cI}) is
Gaussian, because of the central limit theorem. As a consequence,
the probability density function (PDF) of $c_I$ is also Gaussian,
as demonstrated in Appendix B. In fig.~\ref{Fig:PDFBrownian} we
show the PDF of $c_I$ for various $\tau$ for a Brownian suspension
of particles. The symbols are the experimental data, while the
lines are Gaussian PDFs with mean $\overline{c_I(\tau)}$ and
standard deviation $\sigma_{c_I}(\tau)$ obtained directly from the
$c_I$ time series, without any fitting parameters. An excellent
agreement between the data and the theoretical shape of the
distributions is observed at all $\tau$.

\begin{figure}
\includegraphics{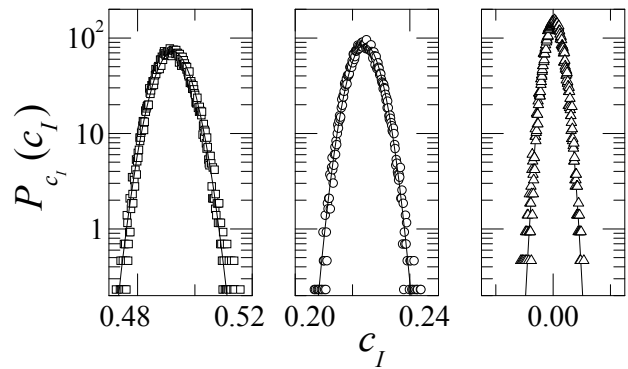}
 \caption{Symbols: PDF of $c_I(t,\tau)$, $P_{c_I}(c_I)$, for
 the same suspension of Brownian particles as for the preceding figures. From left
 to right, the normalized delay $\tau/\tau_0$ is 0.09, 0.87, and
 8.7. In the three panels the $c_I$ axis spans an interval of equal width (0.04). The lines are Gaussian distributions whose mean
 and standard deviations are directly obtained from the $c_I$ time
 series, without any fitting parameter.}
 \label{Fig:PDFBrownian}
\end{figure}

\section{Temporally heterogeneous dynamics: corrections for the noise contribution}
\label{sec:noisedeconvolution}

In temporally heterogeneous dynamics, the fluctuations of $c_I$
are due both to the noise and to dynamical heterogeneity. In this
section we propose a method for correcting the variance and the
PDF of $c_I$ for the noise contribution, so as to obtain the
statistics of the fluctuations due to dynamical heterogeneity.
Moreover, we show that in some instances the full temporal
evolution of $c_I$ may be corrected for the noise, thus allowing
$g_2(t,\tau)-1$ to be determined, not only its variance and PDF.

\subsection{Correction of the variance of $c_I$}
\label{subsec:correctvar}

In the case of dynamically heterogeneous processes, the parameters
$a_1,...,a_k$ in the two-time correlation function
$g_2(a_1(t),...,a_k(t);\tau)-1$, Eq.~(\ref{Eq:cInoise}), fluctuate
with $t$. Therefore, an extra term, $\sigma_{g_2}$, contributes to
the variance of $c_I$, in addition to the noise term analyzed in
the preceding section:
\begin{eqnarray}
   \sigma^{2}_{c_I}(\tau) = N^{-1}\sum_{l=0}^{3}\alpha_l\overline{c_I(\tau)}^{\,l} + \sigma^2_{g_2}(\tau).
   \label{Eq:varcIgeneral}
\end{eqnarray}
In writing the expression above, we have assumed that no
correlation exists between the noise $n(t,\tau)$ due to the finite
number of pixels and the fluctuations of $a_1,...,a_k$. Using
Eq.~(\ref{Eq:var}), $\sigma_{g_2}^2$ may be expressed as
\begin{eqnarray}
   \sigma^{2}_{g_2}(\tau) & = & \sum_{i=1}^{k} \left [ \frac{\partial g_2}{\partial a_i}\right]_{a_i = \overline{a_i}}^{2}
   \sigma_{a_i}^2
   \nonumber \\
   {}&  + & \sum_{i \neq j}  \left [ \frac{\partial g_2}{\partial a_i}\frac{\partial g_2}{\partial
   a_j} \right]_{a_i = \overline{a_i} \atop a_j = \overline{a_j}}\sigma_{a_i,a_j}.
   \label{Eq:varg2}
\end{eqnarray}
An example where $\sigma^{2}_{g_2}$ assumes a particularly simple
form is given by the dynamics of a shaving cream foam, resulting
from intermittent bubble rearrangements, as measured in a DWS
experiment. We find that the fluctuations in the instantaneous
decay rate $\gamma(t)$ of the correlation function are slow
compared to $\overline{\gamma}$, so that at any given time $g_2-1$
is well approximated by a stretched exponential, $\beta
\exp[-(\gamma(t)\tau)^{\mu(t)}]$. Small variations of $\mu(t)$
account for slight changes of the decay rate on a time scale
comparable to $\overline{\gamma}^{\,-1}$. For example, if the
dynamics tend to slow down during the measurement of $g_2-1$, the
initial decay of the correlation function will be faster than its
final decay. Thus, the shape of $g_2$ will be more stretched than
the average one ($\mu < \overline{\mu})$. Conversely, $\mu >
\overline{\mu}$ if the dynamics accelerate during the measurement
of $g_2-1$. By taking into account the fluctuations of both
$\gamma$ and $\mu$, Eq.~(\ref{Eq:varg2}) yields, for the foam,
\begin{eqnarray}
   \sigma^{2}_{g_2}(\tau) & =&  (\overline{\gamma} \tau)^2  \beta^2
   \exp \left[- 2(\overline{\gamma} \tau)^{\overline{\mu}\,} \right ]
   \nonumber \\
   & \times & \left \{ \left ( \frac{\overline{\mu}}{\overline{\gamma}}
   \sigma_{\gamma}\right)^2
    + \left [\, \ln (\overline{\gamma} \tau) \sigma_{\mu} \right]^2
    \right \}
   \label{Eq:varg2foam}
\end{eqnarray}
with $\sigma^{2}_{\gamma} = \overline{(\gamma -
\overline{\gamma})^2}$ and $\sigma^{2}_{\mu} = \overline{(\mu -
\overline{\mu})^2}$. Equation~(\ref{Eq:varg2foam}) extends a
similar expression given in ref.~\cite{MayerPRL2004}, where only
the fluctuations of $\gamma$ where taken into account. However,
here we still neglect possible correlations between $\gamma$ and
$\mu$ that could be described by including the second term of the
r.h.s. of Eq.~(\ref{Eq:varg2}).

A direct test of Eq.~(\ref{Eq:varg2foam}) is not possible, since
we experimentally only access $\sigma^{2}_{c_I}$, not
$\sigma^{2}_{g_2}$. Therefore, for the foam as well as for the
general case of temporally heterogeneous dynamics it is desirable
to subtract the trivial contribution of the measurement noise from
the experimentally measured $\sigma^{2}_{c_I}$, in order to obtain
the physically relevant variance $\sigma^{2}_{g_2}$. We have
tested two different approaches. In the first method, the linear
dependence of $\sigma_{x_i,x_j}$ on $\overline{c_I}$ shown in
fig.~\ref{Fig:sigma} is used to derive formulas for these
quantities that are independent of the instantaneous dynamics and
thus of the homogeneous $vs.$ heterogeneous nature of the
dynamics. These formulas are presented in Appendix A.
Unfortunately, although they provide separately fairly good
estimates of the various terms $\sigma_{x_i,x_j}$, when combined
using eq.~\ref{Eq:varcI1} to evaluate $\sigma^2_{n}$ uncertainties
add up leading to errors of the order of $30-40\%$, as tested on
the data for the single scattering experiment on the Brownian
particles.

\begin{figure}
\includegraphics{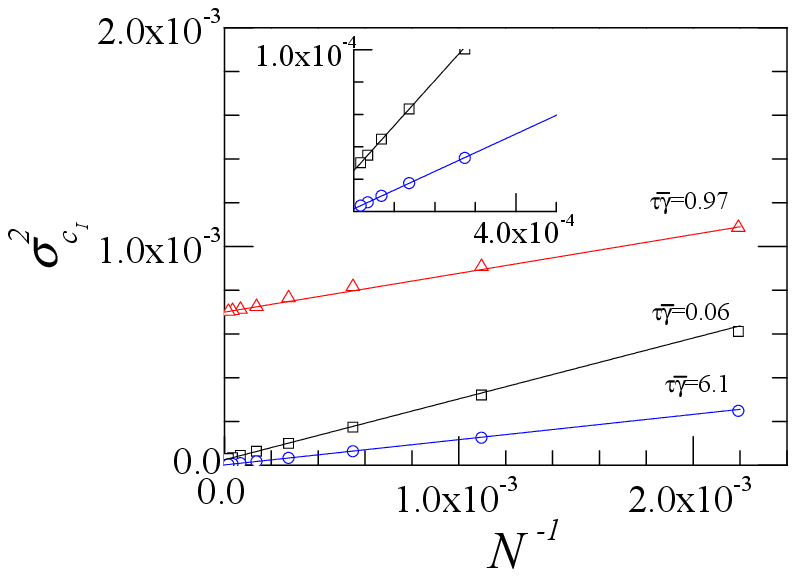}
 \caption{Variance of $c_I(t,\tau)$ for a foam, as a function of the inverse number
 of pixels over which $c_I$ is averaged, for three time delays
 $\tau$. The lines are linear fits to the data for $N^{-1} < 5.5 \times 10^{-4}$.
 For $\tau \overline{\gamma}\approx 1$, the intercept of the fit strongly departs from zero, indicating
 temporarily heterogeneous dynamics as discussed in the text.
 Note that the fit yields a non-zero intercept also at small delay, as shown in the inset that zooms into the small $\sigma_{c_I}^{2}$,
 large $N$ region. Data were collected in the DWS transmission geometry for a duration $T_{\rm exp} = 160$ sec, 242 times longer than the
 relaxation time of $c_I$, $\overline{\gamma}^{-1} = 0.33$ sec.}
 \label{Fig:varcIvsNfoam}
\end{figure}

In the following, we describe in detail a second method that has
proven to be highly effective. The key point is to recognize that,
contrary to the noise term, the fluctuations of
$g_2(a_1,...,a_k;\tau)$ do not depend on the number of pixels over
which $c_I$ is averaged. This is because in the far field geometry
of the scattering experiments described in this work, each CCD
pixel collects light scattered by the whole illuminated sample.
Thus, any spatial or temporal heterogeneity of the dynamics
affects in the same way the signal measured by each pixel. The
different pixel-number dependence of the noise and the
fluctuations (first and second term in the r.h.s. of
Eq.~(\ref{Eq:varcIgeneral}), respectively) suggests a way to
discriminate between these two contributions. We analyze the
speckle images by processing different number of pixels, as
described for the Brownian suspension in
sec.~\ref{SEC:fluctnoise}, and plot $\sigma^{2}_{c_I}(\tau)$ as a
function of $N^{-1}$, as shown in fig.~\ref{Fig:varcIvsNfoam}. As
indicated by Eq.~(\ref{Eq:varcIgeneral}), the slope of a linear
fit to the data yields
$\sum_{l=0}^{3}\alpha_l\overline{c_I(\tau)}^{\,l}$, while the
intercept at $N^{-1} = 0$ is $\sigma_{g_2}^2(\tau)$, the desired
variance of the correlation function due to dynamical
heterogeneity. Thus, the representation of
figs.~\ref{Fig:varcIvsN} and~\ref{Fig:varcIvsNfoam} allows one to
extrapolate $\sigma_{c_I}^{2}$ to $N = \infty$, where the
measurement noise vanishes. As seen in the inset of
fig.~\ref{Fig:varcIvsNfoam}, for $\tau \overline{\gamma}\ll 1$ or
$\tau \overline{\gamma} \gg 1$ the intercept of the linear fit is
very close to zero, indicating that at these delays the
fluctuations of $c_I$ are mainly determined by the measurement
noise. By contrast, at intermediate delays the intercept clearly
departs from zero, revealing the intermittent nature of the
dynamics of the foam.

\begin{figure}
\includegraphics{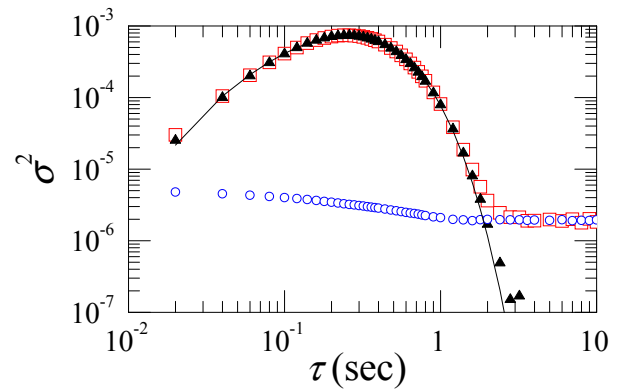}
 \caption{Variance of $c_I(t,\tau)$ for a foam, as a function of $\tau$.
 Open squares: variance of the raw data ($c_I$); open circles: noise contribution;
 solid triangles: $\sigma^{2}_{g_2}(\tau)$, the contribution due to the fluctuations of the number of
 rearrangement events per unit time. The line is a fit to
 $\sigma^{2}_{g_2}(\tau)$ using Eq.~(\ref{Eq:varg2foam}).
 Data were collected in the DWS transmission geometry.}
 \label{Fig:varcIfoam}
\end{figure}

In fig.~\ref{Fig:varcIfoam} we plot the $\tau$ dependence of
$\sigma_{c_I}^{2}$ measured for the foam, together with the noise
contribution, $\sigma_{n}^{2} = N^{-1}
\sum_{l=0}^{3}\alpha_l\overline{c_I(\tau)}^{\,l}$ , and that of
the intrinsic fluctuations, $\sigma_{g_2}^{2}$. The noise
contribution is extracted from the slope of $\sigma_{c_I}^2$ $vs.$
$N^{-1}$, while the variance of the intrinsic fluctuations is
given by the $N^{-1}\rightarrow 0$ limit of $\sigma_{c_I}^{2}$
$vs.$ $N^{-1}$. At all $\tau$ we find $\sigma_{c_I}^{2} =
\sigma_{n}^{2}+\sigma_{g_2}^{2}$ within $0.8 \%$, thus confirming
that our analysis allows to correctly account for the two
contributions to the fluctuations of $c_I$. Note that, while the
shape of the time-averaged correlation function $\overline{g_2} -
1$ is almost the same for the foam and the Brownian particles (a
slightly stretched exponential for the former and a simple
exponential decay for the latter), the $\tau$ dependence of the
fluctuations of $c_I$ is very different (compare the inset of
fig.~\ref{Fig:varcI} to fig.~\ref{Fig:varcIfoam}), thus allowing
temporally heterogeneous dynamics to be unambiguously detected.
For the foam, correcting $\sigma_{c_I}^{2}$ for the noise
contribution is especially important at time delays far from the
mean relaxation time, because the intrinsic fluctuations die off
for $\tau \rightarrow 0$, when virtually no rearrangement had a
chance to occur, and for $\tau \rightarrow \infty$, when so many
rearrangement events occurred that the statistical fluctuations of
their number are negligible. By contrast, we recall that
$\sigma_{n}^{2}$ remains finite at all $\tau$. Once corrected for
the noise contribution, $\sigma_{g_2}^{2}(\tau)$ is very well
described by Eq.~(\ref{Eq:varg2foam}), as shown by the line in
fig.~\ref{Fig:varcIfoam}. Interestingly, the fluctuations are
maximal on the time scale of the mean relaxation time, a general
feature found also in the dynamic susceptibility $\chi_4$ measured
in simulations. Intuitively, this can be explained by recognizing
that the correlation function is most sensitive to a change of the
instantaneous relaxation time for $\tau \approx \tau_0$. Finally,
note that when comparing the absolute values of
$\sigma_{g_2}^{2}(\tau)$ and $\chi_4$ one should keep in mind that
the latter is usually defined as the variance of the correlation
function multiplied by the number $M$ of particles in the system.
Therefore, for homogeneous dynamics $\chi_4 \sim 1$, since the
variance of the number fluctuations is of order $1/M$, while
$\chi_4
> 1$ for heterogeneous dynamics. In the case of the foam shown in
fig.~\ref{Fig:varcIfoam}, taking $M$ as the number of bubbles in
the scattering volume leads to $M \sigma_{g_2}^{2}(\tau) \sim
1000$, indicating strongly heterogeneous dynamics.

\subsection{Correction of the PDF of $c_I$}
\label{subsec:PDFdeconvolution}

In the absence of intrinsic dynamical fluctuations, knowledge of
the average degree of correlation, $\overline{c_I}$, and of the
noise variance, $\sigma_{n}^{2}$, is sufficient to fully determine
the PDF of $c_I$ at fixed lag, because $c_I$ is a Gaussian
variable. Heterogeneous dynamical processes, on the contrary, lead
in general to non-Gaussian distributions of $c_I(t,\tau)$
\cite{LucaJPCM2003,BissigPhysChemComm2003,DuriFNL2005}, whose
shape depends on both the dynamical process and the time delay
$\tau$. Because $c_I(t,\tau)$ is the sum of two uncorrelated
random variables, $g_2(a_1(t),...,a_k(t);\tau)-1$ and the noise
$n(t,\tau)$, the PDF of $c_I$ is the convolution of the
probability distribution of $g_2(a_1(t),...,a_k(t);\tau)-1$ with
that of $n(t,\tau)$ \cite{Frieden2001}:
\begin{eqnarray}
   P_{c_I}(c_I) = P_{g_2-1}(g_2-1) \otimes P_{n}(n),
   \label{Eq:PDFconvolution}
\end{eqnarray}
where $P_x(x)$ denotes the PDF of $x$ and $(f \otimes g) (x)= \int
dx' f(x')g(x-x')$ is the convolution product. In order to recover
the physically relevant PDF of $g_2-1$ from the measured
$P_{c_I}(c_I)$, one may use standard Fourier transform techniques
to deconvolute the experimental data, using $P_{n}(n)$ as the
response function~\cite{NumericalRecipes}: $P_{g_2-1} =
\mathcal{F}^{-1} [\mathcal{F}( P_{c_I}) / \mathcal{F} (P_{n})]$,
where $\mathcal{F}$ and $\mathcal{F}^{-1}$ indicate direct and
inverse Fourier transform, respectively. Unfortunately, this
procedure is very sensitive to noise in the data and leads
typically to unstable solutions exhibiting wide oscillations.
Instead, we use a technique similar to the indirect Fourier
transformation (IFT) method used to process static scattering
data. Details on the IFT method can be found in
\cite{GlatterJAppCryst1977,Glatter2002}, here we simply describe
the main steps of our implementation. We first assume that the
unknown PDF of $g_2-1$ at fixed lag $\tau$ may be written as the
linear superposition of a set of suitable functions $\phi_k$:
\begin{eqnarray}
   P_{g_2-1} = \sum_{k=1}^M\mu_k \phi_k(g_2-1).
   \label{Eq:PDFg2expansion}
\end{eqnarray}
It is convenient to choose $\phi_k(x) = {\rm
rect}(x-x_k;2\Delta)$, where
\begin{eqnarray}
   {\rm rect}(x-x_k;2\Delta) = \left\{ \begin{array} {r@{\quad: \quad}l}
   1 & |x - x_k| \leq \Delta
   \\ 0 & {\rm elsewhere} \end{array} \right.
   \label{Eq:rect}
\end{eqnarray}
is a square pulse of width $2\Delta$ centered on $x_k$. $2\Delta$
is taken to be the width of the bins used to calculate the PDF of
${c_I}$, i.e. the separation between the $c_I$ coordinates of the
$M$ experimental $P_{c_I}$ data. Because the convolution is a
linear transformation, Eqs.~(\ref{Eq:PDFconvolution})
and~(\ref{Eq:PDFg2expansion}) yield the following guess for the
PDF of $c_I$:
\begin{eqnarray}
P_{\rm guess}(c_I) & = & \sum_{k=1}^M\mu_k {\rm
rect}(c_I-c_{I,k};\Delta) \otimes P_{n}(c_I)
\nonumber\\
& = & \sum_{i=k}^M\mu_k \sigma^{2}_{n}\sqrt{\frac {\pi}{2}} \left
[{\rm erf}(x^+) - {\rm erf}(x^-) \right]
\label{Eq:PDFcIexpansion}
\end{eqnarray}
with
\begin{equation}
x^\pm = \frac{1}{\sqrt{2}}\,\sigma^{2}_{n} \left(c_I - c_{I,k} \pm
\Delta \right).
\label{Eq:PDFcIexpansion2}
\end{equation}
Here ${\rm erf}(x) = 2\pi^{-1/2} \int_{0}^{x} \exp(-u^2)du$ is the
error function~\cite{Frieden2001} and $c_{I,k}$ is the center of
the $k$-th bin used to calculate the experimental PDF. In writing
the last line of~Eq.~(\ref{Eq:PDFcIexpansion}) we have used
$P_{n}(c_I) =
(2\pi\sigma^{2}_{n})^{-1\!/2}\exp[-c_I^2/2\sigma^{2}_{n}]$ and
calculated explicitly the convolution product. We note that the
width of the ${\rm rect}$ functions is typically much smaller than
that of the Gaussian noise $P_n$, so that the difference of the
erf functions in square brackets is very close to a Gaussian. We
stress that the only unknowns in Eqs.~(\ref{Eq:PDFcIexpansion})
and (\ref{Eq:PDFcIexpansion2}) are the coefficients $\mu_k$, since
the variance of the noise $\sigma^{2}_{n}$ can be obtained
directly from the experimental $c_I$ using the extrapolation
scheme described in the preceding subsection.

In principle, the coefficients $\mu_k$ can be determined by
fitting Eq.~(\ref{Eq:PDFcIexpansion}) to the experimentally
measured PDF of $c_I$. Once the $\mu_k$'s are known, the desired
PDF of $g_2-1$ can be obtained by using
Eq.~(\ref{Eq:PDFg2expansion}). In practice, two issues must be
addressed when determining the set of $\mu_k$. Firstly, noise in
the experimental $P_{c_I}$ can make the fitting procedure
unstable. It is therefore convenient to smooth the experimental
$P_{c_I}$ before fitting it by Eq.~(\ref{Eq:PDFcIexpansion}). To
this end, $P_{c_I}$ is approximated by a smooth curve, $P_{\rm
sm}$, obtained by fitting the data by a suitable function. We find
that in most cases
\begin{eqnarray}
P_{\rm sm}(c_I) =  A {\rm e}^{ \nu_1(c_I-c_{0}) +
\nu_2(c_I-c_{0})^2  - \nu_3\exp\left[\nu_4(c_I-c_{0})\right]}
 \label{Eq:PDFSmoothed}
\end{eqnarray}
fits well the whole experimental PDF, although fitting $P_{c_I}$
piecewise may be sometimes necessary. Note that the Gumbell PDF
corresponds to the case $\nu_1 = \nu_4 = \alpha >0$, $\nu_2 = 0$,
and $\nu_3 = 1$ \cite{ChamonJChemPhys2004}. The coefficients
$\mu_k$ are then found by fitting $P_{\rm guess}(c_I)$ to $P_{\rm
sm}$, rather than directly to $P_{c_I}$. Secondly, the PDF of
$g_2-1$ determined by inserting the $\mu_k$'s thus obtained in
Eq.~(\ref{Eq:PDFcIexpansion}) often exhibits large oscillations.
This is because the erf functions in Eq.~(\ref{Eq:PDFcIexpansion})
do not form an orthogonal basis, as discussed in ref.
\cite{Glatter2002}. This problem can be solved by adding a
stabilization condition when fitting $P_{\rm sm}$ by $P_{\rm
guess}$. We follow ref.~\cite{Glatter2002} and determine the set
of $\mu_k$ by minimizing the following expression:
\begin{eqnarray}
\sum_{l=1}^L[P_{\rm sm}(c_{I,l})-P_{\rm guess}(c_{I,l})]^2 +
\Lambda \sum_{k=1}^{M-1} (\mu_{k+1} - \mu_k)^2.
 \label{Eq:PDFFitLagrangeMultiplier}
\end{eqnarray}
The first term in the above expression corresponds to the usual
sum of squared deviations between the fitting function ($P_{\rm
guess}$) and the data ($P_{\rm sm}$). The second term assigns a
cost to any large variation between successive $\mu_k$ and thus
tends to suppress all fast oscillation of $P_{\rm guess}$. The
relative weight of the two terms is controlled by the Lagrange
multiplier $\Lambda$, whose optimum value is determined as
described in ref. \cite{Glatter2002}.

\begin{figure}
\includegraphics{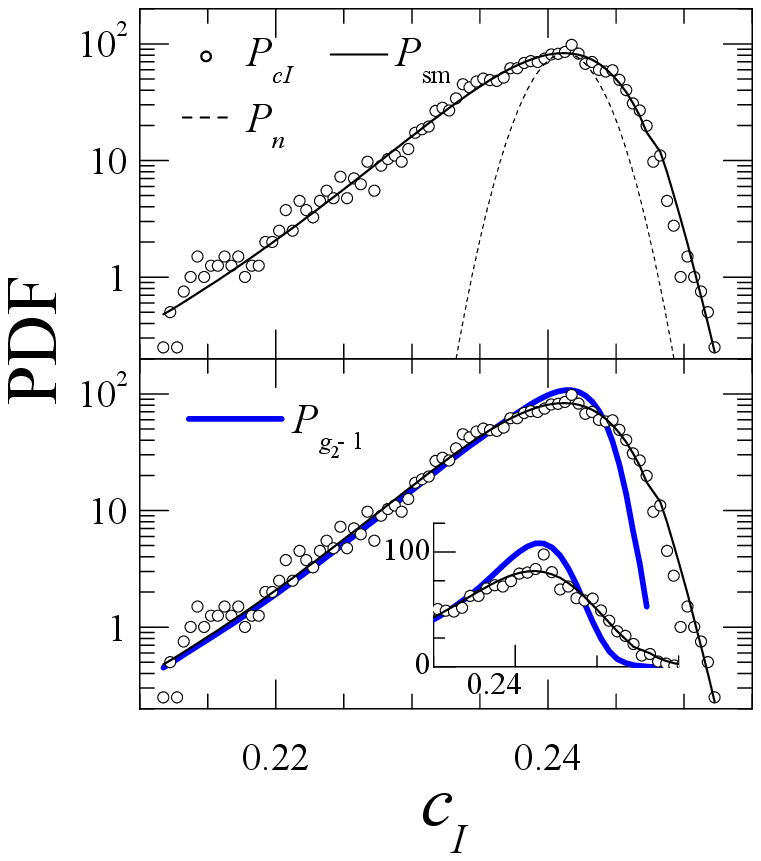}
 \caption{Top panel: PDF of $c_I(t,\tau)$ for a foam, for $\tau = 0.2$ sec (open circles);
 data were collected in the DWS transmission geometry.
 The continuous line is the smoothed distribution, $P_{\rm sm}$, obtained from a fit to Eq.~(\ref{Eq:PDFSmoothed}).
 The dotted
 line is the Gaussian PDF of the noise.
 Bottom panel: the corrected distribution, $P_{g_2-1}$, (thick line), together with the raw data
 and the smoothed PDF (same symbols as in the top panel).
  Note the much steeper decay of the right wing of the corrected PDF.
  The inset shows the data around the peak in a linear axis graph.}
 \label{Fig:PDFfoam}
\end{figure}

The top panel of fig.~\ref{Fig:PDFfoam} shows the PDF of $c_I$
measured in the DWS TRC experiment on foam, for $\tau = 0.02$ sec
(open circles). The solid line is the smoothed PDF, $P_{\rm sm}$,
obtained by fitting the data to Eq.~(\ref{Eq:PDFSmoothed}). The
dotted line is the Gaussian PDF of the noise, $P_n$, whose width
is $\sigma_n$ (for display purposes, $P_n$ has been centered on
$\overline{c_I}$, rather than on 0). In the bottom panel, the
thick line is the PDF corrected for the noise contribution,
$P_{g_2-1}$. Note that the right wing of the corrected PDF drops
much more abruptly than that of the raw data (for comparison, the
uncorrected and the smoothed PDF are also plotted in the bottom
panel). The left wing, on the contrary, is almost unaffected by
the correction. This is a consequence of the nearly exponential
behavior of the left wing: indeed, it can be shown that the
convolution of an exponential function with a Gaussian is again
exponential, with the same growth rate. We test how close the raw
data and the corrected PDF are to a Gumbel distribution by fitting
the data to the expression of Eq.~(\ref{Eq:PDFSmoothed}), with
$\nu_2 = 0$ and $\nu_1 = \nu_3 \nu_4$, corresponding to a
generalized Gumbel PDF \cite{ChamonJChemPhys2004}. For the raw
data, we find $\nu_3 = 1.04$, very close to $\nu_3 = 1$, the value
for a Gumbel PDF. By contrast, for the corrected data $\nu_3 =
0.42$, showing that the right wing of the corrected PDF strongly
departs from both a Gumbel distribution and the ``universal'' PDF
of ref.~\cite{BramwellPRL2000}, for which $\nu_3 = \pi/2 \approx
1.57$. A more detailed investigation of the shape of the PDF of
$c_I$ for various delays $\tau$ and different systems will be
presented elsewhere: here we just stress the importance of the
noise correction in view of any quantitative comparison.

\subsection{Direct correction of $c_I(t,\tau)$ for $\tau << \tau_0$}
\label{subsec:directcorrection}

The methods developed in subsecs.~\ref{subsec:correctvar} and
\ref{subsec:PDFdeconvolution} allow one to calculate the variance
and the PDF of $g_2-1$, but not to correct directly the
time-dependent degree of correlation $c_I(t,\tau)$. Here we show
that such a correction is possible, i.e. that the noise-free
$g_2(t,\tau)-1$ at fixed $\tau$ may be retrieved as a function of
$t$, provided that the following assumptions are fullfilled: $i$)
the dynamics is homogeneous on a time scale comparable to the CCD
exposure time; $ii$) $\tau << \tau_0$, where $\tau_0$ is the
average relaxation time of $g_2-1$.

We first observe that $c_I(t,\tau=0)$ measures the so-called
contrast of the speckle pattern, or coherence factor $\beta$. The
latter is determined only by the speckle-to-pixel size ratio,
which is a time-independent quantity, and by the blurring due to
the fluctuations of the speckle during the time the CCD chip is
exposed \cite{Goodman1975}. If $i$) is fulfilled, the amount of
blurring is constant over time and hence $c_I(t,0)$ fluctuates
only because of the noise \footnote{The so-called Speckle
Visibility Spectroscopy method introduced in refs.
\onlinecite{DixonPRL2003,BandyopadhyayCondmat2005} and discussed
briefly in the conclusions addresses the case where $c_I(t,0)$
fluctuates on the time scale of the CCD exposure time, because of
dynamical heterogeneity.}. Therefore, $n(t,0)$ can be directly
obtained from the experimentally measured $c_I(t,0)$:
\begin{eqnarray}
n(t,0) = c_I(t,0) - \overline{c_I(t,0)}
 \label{Eq:noisetau0}
\end{eqnarray}
If in addition also $ii$) is fulfilled, $n(t,\tau) \approx
n(t,0)$, because the noise evolves on the same time scale as the
speckle pattern, $\tau_0$. Indeed, by analyzing TRC data for
Brownian particles, we have shown in ref.~\cite{DuriFNL2005} that
the noise is highly correlated for $\tau << \tau_0$. This suggests
that $c_I(t,\tau)$ may be directly corrected according to
$g_2(t,\tau)-1  = c_I(t,\tau) - n(t,\tau) \approx c_I(t,\tau) -
n(t,0)$, with $n(t,0)$ obtained via Eq.~(\ref{Eq:noisetau0}). This
approximation may be further refined by taking $n(t,\tau) \approx
[n(t,0)+n(t+\tau,0)]/2$ rather than $n(t,\tau) \approx n(t,0)$ and
by scaling this estimate of the noise so that its standard
deviation matches the actual standard deviation of $n(t,\tau)$:
\begin{eqnarray}
g_2(t,\tau) - 1 & =& c_I(t,\tau) - [n(t,0)+n(t+\tau,0)]/2
\nonumber\\
& \times & \sigma_{n}(\tau)/\sigma_{n}(0).
 \label{Eq:correctcI(t)}
\end{eqnarray}
Here, the standard deviation of the noise of $c_I(t,\tau)$,
$\sigma_{n}(\tau)$, is obtained by applying the extrapolation
scheme described in subsec.~\ref{subsec:correctvar}, while
$n(t,0)$ and $n(t+\tau,0)$ are given by Eq.~(\ref{Eq:noisetau0})
and $\sigma_{n}(0)$ is directly calculated from $n(t,0)$.

\begin{figure}
\includegraphics{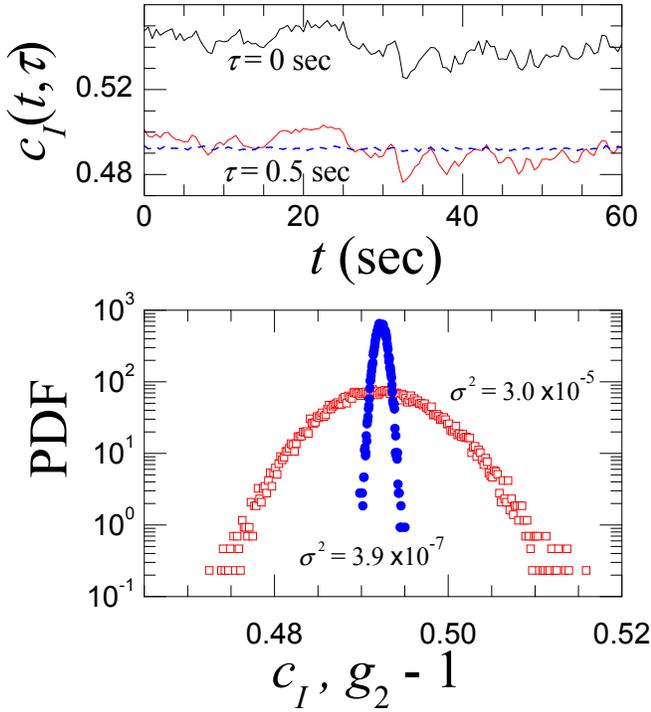}
 \caption{Top panel: $c_I(t,0)$ and $c_I(t,\tau_1=0.5 {\rm ~sec})$ measured for a diluted suspension of Brownian particles, for
 which the relaxation time of $\overline{g_2}-1$ is $\tau_0 = 5.8$. Data are plotted over a short time window to show the strong correlation between the two traces.
 The dashed blue line is $g_2(t,\tau_1)-1$ obtained by correcting $c_I(t,\tau_1)$ for the noise contribution
 according to Eq.~(\ref{Eq:correctcI(t)}):
 the fluctuations due to the noise are almost completely removed.
 Bottom panel: PDF of $c_I(t,\tau_1)$ (open squares) and of $g_2(t,\tau_1)-1$ (solid circles),
 obtained over the full duration of the experiment ($T_{\rm exp} = 10800$ sec). The PDFs are labelled by their variance.}
 \label{Fig:correctioncIBrownian}
\end{figure}

We test Eq.~(\ref{Eq:correctcI(t)}) on TRC data taken for the
dilute suspension of Brownian particles, for which the
fluctuations of $c_I$ are due only to the measurement noise:
$c_I(t,\tau) = n(t,\tau) + \overline{c_I(t,\tau)}$. The top panel
of fig.~\ref{Fig:correctioncIBrownian} shows a portion of
$c_I(t,0)$ and $c_I(t,\tau_1)$, with $\tau_1 = 0.5 {\rm~sec} <<
\tau_0 = 5.8$ sec. Clearly, the two traces are highly correlated,
as expected if $n(t,\tau_1) \approx n(t,0)$. The dotted line is
$g_2(t,\tau_1)-1$ calculated according to
Eq.~(\ref{Eq:correctcI(t)}). Notice that $g_2(t,\tau_1)-1$ is
almost constant, as expected for temporally homogeneous dynamics,
thus demonstrating the effectiveness of the correction. The bottom
panel shows the PDF of $c_I(t,\tau_1)$ and $g_2(t,\tau_1)-1$
calculated over the whole duration of the experiment. By
correcting $c_I(t,\tau_1)$ for the noise, its variance is reduced
by almost a factor of 100 (from $3.0 \times 10^{-5}$ to $3.9
\times 10^{-7}$). The residual fluctuations of $g_2(t,\tau_1)$ are
most likely due to the noise $\epsilon_p(t)$ discussed in
reference to Eq.~(\ref{Eq:Icorrected}), whose variance we estimate
to be of the order of $3.9 \times 10^{-7}$ \footnote{To evaluate
$\sigma^2_{\epsilon}$, we take a time series of dark images. To
each dark image we then add, pixel-by-pixel, the intensity
distribution of one single image of the speckle pattern scattered
by the sample. In the series thus obtained, all images are
identical, except for the small fluctuations due to
$\epsilon_p(t)$. This corresponds to what would be obtained for a
perfectly frozen scatter, in the absence of all possible
experimental artifacts, as discussed in sec. \ref{SEC:artifacts}.
We process the data as usually and calculate the variance of
$c_I$. For all lags $\tau
>0$ the same value is obtained, which is taken as an estimate of
$\sigma^2_{\epsilon}$.}.

\subsection{Influence of the duration of
the experiment} \label{subsec:expduration}

The analysis of the fluctuations of $c_I$ presented in subsecs.
\ref{subsec:correctvar} and \ref{subsec:PDFdeconvolution} was
developed under the assumption that the dynamics be stationary, so
that data could be collected over a period, $T_{\rm exp}$, much
longer than the average relaxation time of the intensity
correlation function, $\tau_0$. Dynamical heterogeneity, however,
appear to be more prominent for systems close to jamming or
quenched in an out-of-equilibrium state. For these systems,
meeting the condition $T_{\rm exp} \gg \tau_0$ is often
impossible, since the sample may be aging, leading to
non-stationary dynamics, or because, even if the dynamics is
stationary, the relaxation time may be as large as several tens of
hours. It is therefore important to address the issue of the
influence of the experiment duration on the measured
$\sigma_{c_I}$.

\begin{figure}
\includegraphics{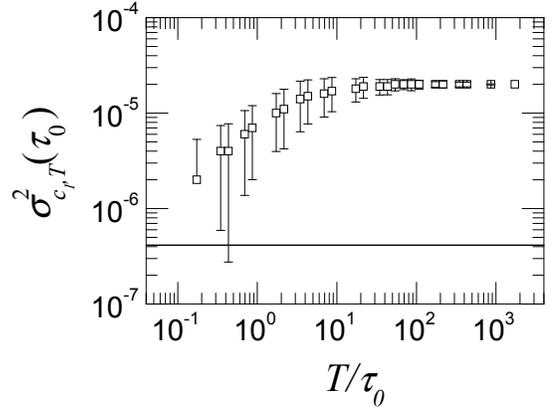}
 \caption{Variance of $c_I(t,\tau = \tau_0)$ calculated over a duration $T$ as a function of
 $T$ normalized by the mean relaxation time $\tau_0$, for a suspension of Brownian particles. A similar behavior
 is observed for all $\tau$. The line is the contribution to the variance of $c_I(t,\tau = \tau_0)$ of the CCD
 electronic noise (see the text for more details).}
 \label{Fig:varcIsegment}
\end{figure}

Let us first consider the simpler case of homogeneous dynamics. We
divide the time series of $c_I(t,\tau)$ obtained for the Brownian
particles into non-overlapping segments of duration $T < T_{\rm
exp}$. We denote by $\sigma^{2}_{c_I,T}(\tau)$ the mean value of
the variance of $c_I(t,\tau)$ calculated for each segment of
duration $T$, and plot in fig.~\ref{Fig:varcIsegment}
$\sigma^{2}_{c_I,T}$ as a function of $T$ normalized by the
relaxation time $\tau_{0}$ (the data refer to $\tau = \tau_0$, a
similar behavior is observed for all $\tau$). For $T/\tau_0
>> 1$, $\sigma^{2}_{c_I,T}$ is independent of $T$, because in this
regime the experiment duration is long enough for the system (and
thus the speckle pattern) to sample a sufficiently large number of
different configurations. Consequently, $\sigma^{2}_{c_I,T}$
saturates to the maximum value given by Eq.~(\ref{Eq:varpoly}),
which is dictated only by $\tau$ and the number of CCD pixels.
Note that for $T/\tau_0$ close to one ($T/\tau_0 = 0.87$),
$\sigma^{2}_{c_I,T}$ is still significantly lower than its
saturation value (about $35\%$), while it increases to $95 \%$ for
$T/\tau_0 > 20$ and saturates only for $T/\tau_0 \geq 50$. As
$T/\tau_0$ decreases below 1, the amplitude of the fluctuations is
significantly reduced when decreasing $T$, since the system is not
given enough time to explore significantly different
configurations. For $T/\tau_0 \ll 1$, the speckle pattern is
essentially frozen on the time scale of the experiment duration:
the fluctuations of $c_I$ due to the evolution of the speckle
pattern are thus expected to be almost completely suppressed. For
the data shown in fig.~\ref{Fig:varcIsegment}, this regime is not
quite reached, since the CCD acquisition rate could not be fast
enough to allow for several images to be acquired on a time scale
much smaller than $\tau_0$ (the smallest lag between images is 0.5
sec = $\tau_0/11.6$). In the $T/\tau_0 \rightarrow 0$ regime, the
main contribution to $\sigma^{2}_{c_I,T}$ should come from the CCD
electronic noise $\epsilon_p(t)$, whose variance,
$\sigma^{2}_{\epsilon}$, is shown as a line in
fig.~\ref{Fig:varcIsegment}. The data shown in this figure clearly
demonstrate that care must be taken when comparing the
fluctuations measured in experiments whose relative duration
$T_{\rm exp}/\tau_0$ is different.

\begin{figure}
\includegraphics{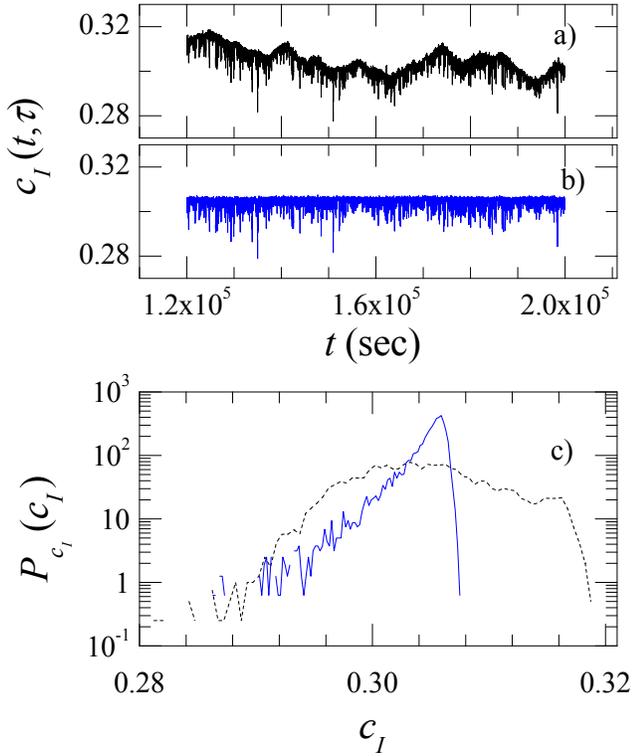}
 \caption{a): time variation of $c_I$ for a multilamellar vesicle gel, for $\tau = 100$ sec.
 The mean relaxation time is $\tau_0 = 5050$ sec, the duration of the experiment is $T = 80000$ sec = $15.8 \times \tau_0$.
  a): raw data; b): the same data after applying the correction  Eq.~(\ref{Eq:correctcI(t)}); c): PDF of the traces shown in a)
  (dotted line) and b) (solid line).}
 \label{Fig:cIonions}
\end{figure}

For heterogeneous dynamics, we expect the behavior of
$\sigma^{2}_{c_I,T}$ to be qualitatively similar, although the
situation will be in general more complicated. In fact, in this
case the relevant time scales to which $T_{\rm exp}$ has to be
compared are not only the mean relaxation time of $g_2-1$,
$\tau_{0}$, but also the characteristic time of the ``intrinsic''
fluctuations of the dynamics, described by the variation of the
parameters $a_1(t),...a_k(t)$ in Eq.~(\ref{Eq:cInoise}). As an
example, for the foam data shown in figs.~\ref{Fig:varcIvsNfoam}
and~\ref{Fig:varcIfoam} $\tau_0 = \overline{\gamma(t)}^{~-1} =
0.32$ sec, while the temporal fluctuations of the decay rate occur
on a much longer time scale, $\tau_{\rm fluct} \approx 7$
sec~\cite{BissigPhD2004} (see also fig. \ref{Fig:seccorfoam} below
and the associated discussion). Hence, in order to measure
precisely the fluctuations the experiment should last more than
about $20 \tau_{\rm fluct}$, rather than just $20 \tau_0$. For
other jammed materials, on the contrary, the intrinsic
fluctuations may be much faster than $\tau_0$, as demonstrated by
the sudden sharp drops of $c_I$ resulting from intermittent
rearrangements in a closely packed multilamellar vesicle
system~\cite{Castro-RomanPRL1999}, shown in
fig.~\ref{Fig:cIonions}a. In this case, at least for the smallest
time lags, the analysis proposed in subsec.
\ref{subsec:directcorrection} allows the slow fluctuations due to
the measurement noise to be suppressed, while preserving the fast
drops of $c_I$ that contain the physically relevant information on
the dynamical intermittency, as seen in fig.~\ref{Fig:cIonions}b.
Note the dramatic change of the shape of the PDF of $c_I$ before
and after correcting the data, as shown in
fig.~\ref{Fig:cIonions}c.

\section{The second correlation}
\label{SEC:seccorr} In the previous sections, the fluctuations of
$c_I$ were analyzed in terms of their variance and PDF. Additional
information on the system dynamics can be obtained by studying not
only the probability distribution of the fluctuations and its
second moment, but also the way these fluctuations occur in time.
We characterize the temporal properties of $c_I$ at a fixed lag
$\tau$ by introducing the time autocorrelation function of
$c_I(t,\tau)$:
\begin{eqnarray}
   C_{c_I}(\Delta t,\tau) = \frac{ \overline{c_I(t,\tau)c_I(t+\Delta t,\tau)} - \overline{c_I}^2 }
   {\overline{c_I(t,\tau)^2} - \overline{c_I}^2}.
   \label{Eq:secondcorrelation}
\end{eqnarray}
With this choice of the normalization, $C_{c_I}(0,\tau) = 1$, and
$C_{c_I}(\Delta t,\tau) = 0$ if $c_I(t,\tau)$ and $c_I(t+\Delta
t,\tau)$ are uncorrelated (e.g. for $\Delta t \rightarrow
\infty$). Because $C_{c_I}$ is the correlation function of a
time-varying quantity --$c_I(t,\tau)$-- which is obtained itself
by correlating the scattered intensity, we shall term it the
``second correlation'', in analogy with the ``second spectrum''
first introduced in the context of spin
glasses~\cite{WeissmanRevModPhys1993}. The second spectrum
describes in the frequency domain the fluctuations around the mean
value of the spectrum of a time-dependent quantity. Similarly, the
second correlation describes ---in the time domain--- the
fluctuations of the degree of correlation between the
time-dependent system configurations. The second correlation is
also similar to the 4-th order intensity correlation function
introduced in refs. \cite{LemieuxJOSAA1999,LemieuxAPPOPT2001},
$g^{(4)}_T (\tau) =
\overline{I(0)I(T)I(\tau)I(\tau+T)}/\overline{I}^{4}$. Note
however that $g^{(4)}_T (\tau)$ compares the scattered intensity,
and thus the sample configuration, at four successive times, while
the second correlation compares the $change$ in sample
configuration occurring over two time intervals of duration $\tau$
separated by a time $\Delta t$.

\begin{figure}
\includegraphics{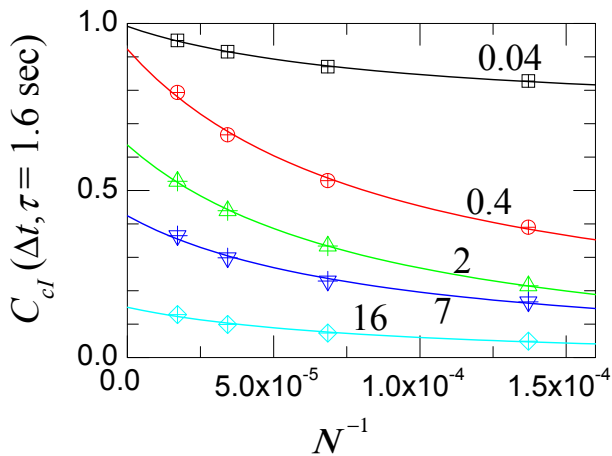}
 \caption{Correction method for $C_{g_2}$. Symbols: the second correlation $C_{c_I}$ as a function of the
inverse number of processed pixels, for various $\Delta t$ (shown
by the labels, in sec) and for $\tau = 1.6$ sec (only the data for
the largest $N$ are shown). The data are obtained from a DWS
measurement on a foam. The lines are fits to the data using
Eq.~(\ref{Eq:secondcorrelationN}); the full range of the fit is
shown in the figure. The outcome of the fit are the values of
$C_{g_2}(\Delta t,\tau)$ and $C_n(\Delta t,\tau)$ shown with the
same symbols in fig.~\ref{Fig:seccorfoam}.}
 \label{Fig:seccorcorrection}
\end{figure}

As for the case of the PDF discussed in
sec.~\ref{sec:noisedeconvolution}, the second correlation contains
contributions from both the physically relevant intrinsic
fluctuations of $g_2(a_1(t),...,a_k(t);\tau)-1$ and the noise,
$n(t,\tau)$. We focus on stationary dynamics measured over a
period $T_{\rm exp} \gg \tau_0$ and assume that the fluctuations
of $g_2(a_1(t),...,a_k(t);\tau)-1$ and $n(t,\tau)$ are
uncorrelated: $\overline{g_2 n} = \overline{g_2} \, \overline{n}$
for all $\Delta t$. Under these assumptions and by using
Eq.~(\ref{Eq:cInoise}), Eq.~(\ref{Eq:secondcorrelation}) yields
\begin{eqnarray}
   C_{c_I}(\Delta t,\tau) = \frac{ \sigma^{2}_{g_2}(\tau) C_{g_2}(\Delta t, \tau) + \sigma^{2}_{n}(\tau) C_n(\Delta t,\tau)}
   {\sigma^{2}_{g_2}(\tau) + \sigma^{2}_{n}(\tau)}~,
   \label{Eq:secondcorrelation2}
\end{eqnarray}
where $C_{g_2}$ and $C_n$ are the correlation functions of the
fluctuations of $g_2$ and $n$, respectively, defined similarly to
Eq.~(\ref{Eq:secondcorrelation}). Experiments on Brownian
particles (homogeneous dynamics) show that $C_n(\Delta t,\tau)
\simeq \overline{c_I}(\Delta t)/\beta$ for all
$\tau$~\cite{DuriFNL2005}. Indeed, $C_n$ is expected to be
proportional to $\overline{c_I}$ because the $\tau$ dependence of
the noise time autocorrelation stems from the same physical
mechanism leading to the decay of the intensity correlation
function, i.e. the renewal of the speckle pattern over time
discussed in sec.~\ref{SEC:fluctnoise}. In order to extract the
desired second correlation of $g_2$ from the noise-affected
$C_{c_I}$, we use a method similar to the extrapolation technique
adopted for the calculation of the variance of $g_2$. At $\Delta
t$ and $\tau$ fixed, $C_{c_I}$ depends on the number $N$ of
processed pixels through the $\sigma^{2}_{n}$ term in
Eq.~(\ref{Eq:secondcorrelation2}). We omit for simplicity the
explicit dependence on $\Delta t$ and $\tau$ and insert
Eq.~(\ref{Eq:varpoly}) into Eq.~(\ref{Eq:secondcorrelation2}):
\begin{eqnarray}
   C_{c_I}(N) = \frac{ \sigma^{2}_{g_2} C_{g_2} + C_n N^{-1}\sum_{l=0}^{3}\alpha_l\overline{c_I}^l }
   {\sigma^{2}_{g_2} + N^{-1}\sum_{l=0}^{3}\alpha_l\overline{c_I}^l}
   \label{Eq:secondcorrelationN}
\end{eqnarray}
The l.h.s. of this expression can be calculated by processing ROIs
of the speckle images of different sizes, as explained in secs.
\ref{SEC:fluctnoise} and \ref{sec:noisedeconvolution}, and plotted
as a function of $N^{-1}$, as shown in
fig.~\ref{Fig:seccorcorrection} for a foam. The r.h.s. of
Eq.~(\ref{Eq:secondcorrelationN}) is then used as a fitting
function for $C_{c_I}(N)$, where the fitting parameters are the
desired $C_{g_2}$ and $C_n$, while $\sigma^{2}_{g_2}$ and
$\sum_{l=0}^{3}\alpha_l\overline{c_I}^l$ are obtained
independently from the correction of the variance of $c_I$, as
explained in subsec. \ref{subsec:correctvar}.

\begin{figure}
\includegraphics{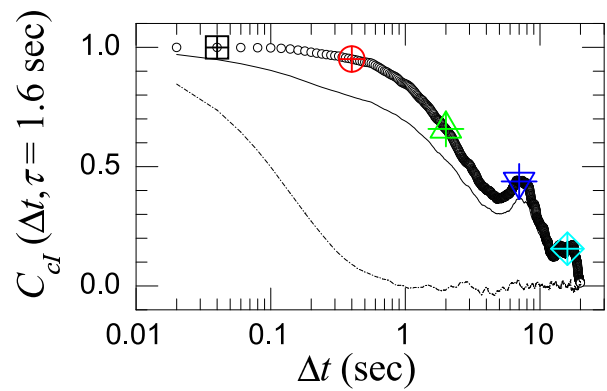}
 \caption{Time autocorrelation function of the fluctuations of $g_2$ for a foam, obtained from the second correlation
 $C_{c_I}(\Delta t,\tau)$ by applying the correcting scheme described in the text. Small open circles: the noise-free $C_{g_2-1}$;
 solid line: the experimentally measured $C_{c_I}$; dashed line: the noise contribution, $C_n$. The large symbols
 are the values of $C_{g_2-1}$ obtained from the fit to the corresponding data shown in fig.~\ref{Fig:seccorcorrection}.
 Note the peaks of $C_{g_2-1}$
 at $\Delta t \approx 7$ sec, and $\Delta t \approx 14$ sec,
 suggesting pseudoperiodic rearrangement events, as discussed in the text.}
 \label{Fig:seccorfoam}
\end{figure}

By repeating this procedure for all $\Delta t$ and $\tau$ of
interest, the full second correlation can be corrected for the
noise contribution. We show in fig.~\ref{Fig:seccorfoam}
$C_{g_2}(\Delta t, \tau)$ for a foam (open circles), together with
the uncorrected data ($C_{c_I}$, solid line) and the noise
contribution ($C_{n}$, dashed line).
Remarkably, a peak is visible in $C_{g_2}$, at $\Delta t = \Delta
t^* \approx 7$ sec. Note that the peak is not present in $C_n$,
whose only relevant time scale is that of the relaxation of
$\overline{c_I}$, $\overline{\gamma}^{-1} = 0.32$ sec. Therefore,
the peak in $C_{g_2}$ must be associated with the intrinsic
fluctuations of the dynamics due to the intermittent bubble
rearrangements. Indeed, the peak indicates that the fluctuations
of the instantaneous decay rate $\gamma(t)$ are pseudoperiodic on
a time scale of the order of $\Delta t^*$. We are currently
investigating the origin of this feature.

\section{Possible artifacts}
\label{SEC:artifacts}

\begin{figure}
\includegraphics{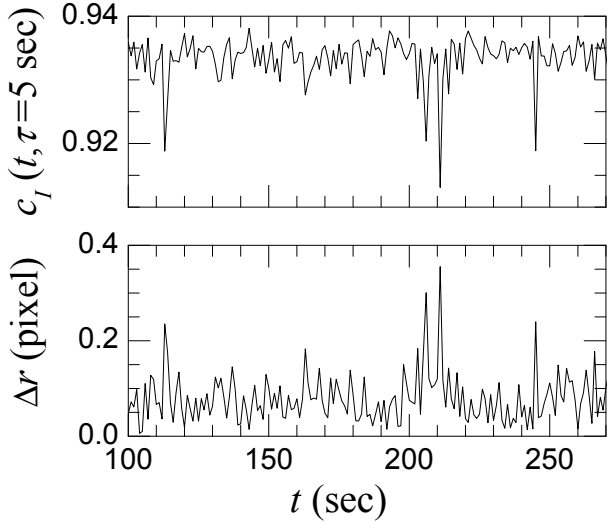}
 \caption{Top panel: $c_I(t,\tau)$ measured for a static scatterer, a ground glass, for $\tau = 5$ sec. Note the sharp drops
 of the degree of correlation that greatly exceed the noise of the data. Bottom panel: rigid displacement between
 the pair of speckle images used to calculate $c_I(t,\tau)$ above, measured by PIV. The sharp drops of $c_I$ are artifacts
 due to
 larger-than-average rigid shifts of the speckle images.}
 \label{Fig:groundglassspikes}
\end{figure}

In a TRC measurement, the focus is on the fluctuations of the
degree of correlation, rather than on its mean value. As a
consequence, TRC measurements are very sensitive to various
sources of spurious fluctuations, whose effects are usually less
prominent or even negligible in traditional dynamic light
scattering experiments, since they tend to average out. An example
is provided in the top panel of fig.~\ref{Fig:groundglassspikes},
which shows $c_I$ measured for the speckle pattern generated in
the transmission geometry by a ground glass, a scatterer whose
dynamics are completely frozen-in. In this case, we would expect
$c_I$ to fluctuate slightly only because of the electronic noise
$\epsilon_p(t)$ discussed in sec.~\ref{SEC:measure}. Surprisingly,
sharp drops of the degree of correlation are clearly visible.
Because the drops are rare, they do not affect significantly
$\overline{c_I}$; however, they do change significantly the PDF of
$c_I$ and the second correlation. Indeed, one would mistakenly
take the dynamics to be intermittent, if the sample was not known.
The origin of this artifact is the laser beam pointing
instability. Beam pointing instability is the characteristic noise
of lasers that results in small fluctuations of the propagation
direction of the output beam. Since the speckle pattern is
centered around the propagation direction, any change of the
incoming beam direction entails a rigid shift of the speckle image
with respect to the CCD detector. Thus, the intensity at each
pixel slightly changes, leading to a drop of $c_I$. To demonstrate
that this mechanism is indeed responsible for the sharp drops of
$c_I$ observed in fig.~\ref{Fig:groundglassspikes}, we show in the
bottom panel the corresponding shift, $\Delta r(t,\tau)$, between
pairs of images taken at time $t$ and $t+\tau$, measured by
Particle Imaging Velocimetry (PIV) \cite{TokumaruExpInFluids1995}.
This method is based on spatial cross-correlation techniques and
allows the rigid motion between two images to be quantified with
sub-pixel resolution (our adaption of PIV to speckle imaging will
be described elsewhere). A comparison between the two panels of
fig.~\ref{Fig:groundglassspikes} clearly shows that the
anomalously large drops of $c_I$ are due to larger-than-average
rigid shifts of the speckle images. Note that shifts as small as a
fraction of pixel, corresponding to a few microns, have a
measurable impact on $c_I$.

This example shows how sensitive to instabilities TRC is.
Therefore, care must be taken in order to minimize possible
instabilities and to identify any spurious fluctuations of $c_I$.
Moreover, attempts should be done to correct $c_I$ for artifacts.
In our experience, the most common artifact encountered in TRC
experiments is similar to that exemplified by
fig.~\ref{Fig:groundglassspikes}: fluctuations or sharp drops of
$c_I$ due to a rigid shift of the speckle pattern, rather than to
the characteristic ``boiling'' or flickering of the speckle images
associated with the evolution of the sample configuration. In
addition to beam pointing instability, temperature variations are
often to blame for spurious features in the temporal evolution of
$c_I$.

Temperature variations may change the direction of propagation of
light because of refractive effects, since the refractive index,
$n_{\rm D}$, varies with temperature, $T$. As an example, let us
consider a typical small-angle single scattering measurement on an
aqueous sample contained in a rectangular cell, whose entrance and
exit walls are perpendicular to the incoming beam. Light scattered
at an angle $\theta_{\rm s}$ equal to, e.g., 15 deg is refracted
at the water-to-air interface and exits at an angle given by
Snell's law: $\theta_{\rm air} = \arcsin \left(n_{\rm
D}\sin\theta_{\rm s}\right )$~\footnote{We neglect refractions at
the solvent-container and container-air interfaces, since they
only displace the beam parallel to itself.}. A temperature
fluctuation $\delta T$ induces a variation $\delta \theta_{\rm
air}$ given by
\begin{eqnarray}
    \delta \theta_{\rm air}  & = & \frac{\partial \theta_{\rm air}}{\partial n_{\rm D}} \frac{\partial n_{\rm
    D}}{\partial T} \delta T
    \nonumber\\
     & = & \frac{\sin \theta_{\rm s} \partial n_{\rm D}/ \partial T}{ \sqrt{1 - \sin^2 \theta_{\rm s}n_{\rm
     D}^2}} \, \delta T \approx 2.8 \times 10^{-5} {\rm~rad},
    \label{Eq:dthetaair}
\end{eqnarray}
where we have used $n_{\rm D}=1.33$, $\frac{\partial n_{\rm
D}}{\partial T} \approx 10^{-4}~{\rm K}^{-1}$ for water and we
have taken $\delta T = 1$ K. In a typical small angle setup the
sample-to-CCD distance is at least 10 cm and the pixel size is
about 10 $\mu$m, resulting in a shift of the speckle associated to
the direction $\theta_{\rm s}$ of the order of 0.28 pixels,
sufficient to significantly reduce $c_I$. Thus temperature
fluctuations of the order of 1 K may lead to measurable spurious
fluctuations of $c_I$ because of refractive effects.

In most single scattering wide-angle apparatuses the sample cell
is cylindrical and both the incoming beam and the scattered light
cross all optical interfaces at normal incidence, so that
refractive effects are avoided. Nevertheless, any change of the
refractive index due to temperature variations would still result
in a change of the speckle pattern. This is because each speckle
is associated to a well defined value of the scattering vector,
$\textbf{\textrm{q}}_{\rm s}$. If $n_D$ changes, the scattering
angle $\theta_{\rm s}$ corresponding to $\textbf{\textrm{q}}_{\rm
s}$ is modified, since $\theta_{\rm s} = 2\arcsin
\left(\frac{\lambda_0 q_{\rm s} }{4\pi n_{\rm D}} \right )$, where
$\lambda_0$ is the laser in-vacuo wavelength. Accordingly, the
speckle pattern is contracted or dilated around the $q = 0$
direction. This is the same effect as the radial shift of the
speckle pattern when changing the wavelength of the incident
radiation, which was studied in the
Seventies~\cite{ParryOptCommun1974}. When only a limited portion
of the speckle pattern corresponding to a small solid angle is
imaged on the CCD detector (as it is the case, e.g., for the
single scattering experiments at $\theta = 45$ deg reported in
this paper), such a global contraction or dilation results,
locally, in a rigid shift of the speckles. The change $\delta
\theta_{\rm s}$ of $\theta_{\rm s}$ in response to a temperature
fluctuation $\delta T$ is
\begin{eqnarray}
    \delta \theta_{\rm s}   =  \frac{\partial \theta_{\rm s}}{\partial n_{\rm D}} \frac{\partial n_{\rm
    D}}{\partial T} \delta T
    \nonumber\\
      =  \frac{-2 \lambda_0 q_{\rm s} \,\partial n_{\rm D}/ \partial T}{n_{\rm
     D} \sqrt{16 \pi^2 n_{\rm
     D}^2 + \lambda_0^2q_{\rm s}^2}} \,\delta T \approx -6.2 \times 10^{-5} {\rm~rad},
    \label{Eq:dthetas}
\end{eqnarray}
where we have used $n_{\rm D}=1.33$, $\lambda_0 = 0.535$ $\mu{\rm
m}$, $q_{\rm s} = 12.44$ $\mu{\rm m}^{-1}$ ($\theta_s = 45$ deg),
$\frac{\partial n_{\rm D}}{\partial T} \approx 10^{-4}~{\rm
K}^{-1}$ and $\delta T = 1$ K. Taking the sample-to-CCD distance
to be 10 cm, we find that a fluctuation $\delta T = 1$ K would
result in a shift of the speckle pattern of 6.2 $\mu{\rm m}$,
comparable to the pixel size. This would lead to a catastrophic
drop of $c_I$, since the speckle size is typically of the order of
the pixel size. Indeed, even a fluctuation ten time smaller,
$\delta T = 0.1$ K, would have a measurable impact on $c_I$.
Similar arguments apply also to multiple scattering experiments.
Note that DWS experiments are more sensitive to variations of
$n_D$ than single scattering measurements are, because minute
changes of $q$ at each scattering event add up along the photon
path, eventually resulting in a significant change of the phase of
scattered photons.


Various strategies are possible to avoid the artifacts discussed
above, or at least to mitigate their effects on TRC data. The
impact of beam pointing instability can be minimized by reducing
as much as possible the light path between the laser and the
sample, or by delivering the beam via a fiber optics. In the
latter case, beam pointing instability results in fluctuations of
the laser-to-fiber coupling efficiency and thus of the incident
intensity, $I_{\rm in}$. Because $c_I$ is normalized with respect
to the instantaneous pixel-averaged intensity (see
Eq.~\ref{Eq:cI}), fluctuations of $I_{\rm in}$ have little if any
effect. Temperature should be controlled at least to within 0.1 K
and the sample or sample holder temperature should be monitored,
so that any suspect feature in $c_I$ could be compared to the
temperature record. A PIV analysis similar to that presented in
fig.~\ref{Fig:groundglassspikes} is a useful test to check whether
a spurious rigid shift of the speckle pattern is at the origin of
large drops of $c_I$.

The input from PIV measurements may also be used to correct for
the effect of a rigid shift $\Delta \textbf{\textrm{r}}$ of the
speckle pattern. This could be achieved by constructing a
corrected speckle image, shifting back by $-\Delta
\textbf{\textrm{r}}$ the second image of the pair used to compute
$c_I$. The corrected image would then be used to calculate $c_I$.
Standard image-processing techniques \cite{LehmannIEEE1999} can be
used to obtain sub-pixel shifts. Alternatively, one may exploit
the fact that in the presence of both a rigid shift $\Delta
\textbf{\textrm{r}}$ and a genuine evolution of the speckle
pattern, the measured degree of correlation, $c_{I,{\rm meas}}$,
factorizes as $c_{I,{\rm meas}}(t,\tau) =
c_I(t,\tau)\left[g_{2,{\rm sp}}(\Delta
\textbf{\textrm{r}})-1\right]$ (a proof is given in the Appendix
of ref.~\cite{PhamRSI2004}). Here $g_{2,{\rm sp}}(\Delta
\textbf{\textrm{r}})-1$ is the normalized spatial autocorrelation
function of the speckle image and $c_I(t,\tau)$ is the degree of
correlation that would be measured in the absence of any shift. We
are currently exploring both approaches.

\begin{figure}
\includegraphics{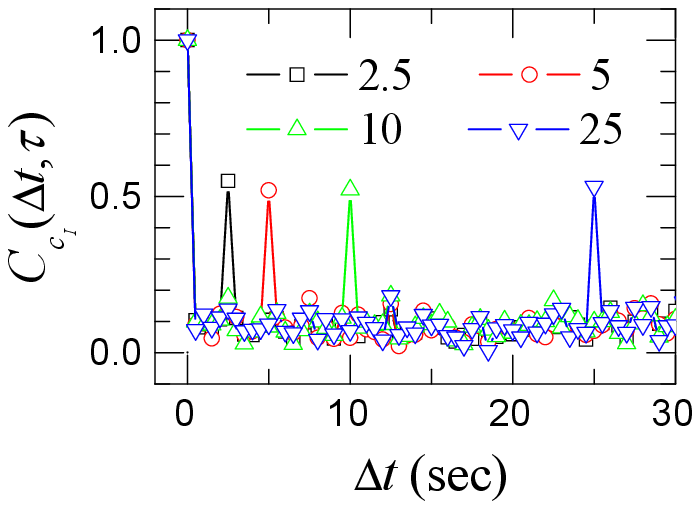}
 \caption{Second correlation measured for a perfectly frozen speckle pattern. The curves are labelled by
 $\tau$, in sec. Note the spurious peaks at $\Delta t = \tau$, due to the CCD dark noise, as
 discussed in the text.}
 \label{Fig:SecCorrCCDNoise}
\end{figure}

We conclude this section by reporting a spurious feature due to
the electronic noise $\epsilon_p$, which may affect the second
correlation $C_{c_I}(\Delta t,\tau)$ measured for time lags $\tau$
much smaller than the typical relaxation time. On those short time
scales, the speckle pattern is essentially frozen; therefore, we
shal use as an example the series of images of a perfectly frozen
speckle pattern obtained as described at the end of subsec.
\ref{subsec:directcorrection}. We process the data as usually and
calculate the variance of $c_I$. For all lags $\tau
>0$ the same value is obtained, which is taken as an estimate of
$\sigma^2_{\epsilon}$. This has the advantage of excluding other
artifacts, such as a possible rigid shift of the speckle pattern.
As shown in fig.~\ref{Fig:SecCorrCCDNoise}, the second correlation
exhibits spikes at $\Delta t = \tau$, emerging form a baseline
close to zero. This is a spurious effect due to the fact that the
electronic noise $\epsilon_p$ is delta-correlated in time.
Therefore, $\epsilon_p$ contributes an extra term to
$\overline{c_I(t,\tau)c_I(t+\Delta t,\tau)}$ whenever two out of
the four intensity terms involved in this expression are measured
at the same time, as for $\Delta t = \tau$. A rigorous calculation
can be found in ref.~\cite{BissigPhD2004}, showing that the height
of the spurious peaks is 0.5, in good agreement with
fig.~\ref{Fig:SecCorrCCDNoise}. The spikes of the second
correlation shown in a figure of a (withdrawn) preprint authored
by some of us \cite{BissigCondmat2003} are most likely due to this
effect.

\section{Conclusions}
\label{SEC:Conc} We have shown that TRC allows heterogeneous
dynamics to be unambiguously discriminated from homogeneous
dynamics. In order to quantify dynamical heterogeneity, three
statistical objects are particularly insightful: the variance of
$c_I$, which corresponds to the dynamical susceptibility $\chi_4$,
the PDF of the degree of correlation, and its time autocorrelation
function --the second correlation. Statistical noise due to the
finite number of pixels over which $c_I$ is averaged can
contribute significantly to the fluctuations of the degree of
correlation. Thanks to the correction scheme described in this
paper, the variance, PDF, and time autocorrelation of $c_I$ can be
corrected for this contribution, thus making quantitative
measurements of heterogeneous dynamics accessible to scattering
techniques. These corrections are particularly important when the
intrinsic fluctuations are comparable to or even smaller than the
noise. This may occur because dynamic heterogeneity is mild, e.g.
in moderately concentrated suspensions of colloids
\cite{BallestaInPreparation}, or because the number of available
pixels is reduced, e.g. in a small angle light scattering or XPCS
setup, where the lowest accessible $q$ vectors correspond to small
rings centered around the transmitted beam and containing a
limited number of pixels \cite{LucaRSI1999}. Corrections are also
important when comparing TRC data obtained from different
apparatuses, for which both the speckle size and the number of
pixels may differ, or when analyzing small angle data, since the
number of pixels in the rings associated to different $q$ vectors
varies with $q$.

In addition to providing a quantitative description of dynamical
fluctuations for systems in a stationary or quasi-stationary
state, TRC is a useful tool for studying rapidly evolving
dynamics, e.g. during gelation \cite{BissigPhD2004}, as well as
the response to an instantaneous perturbation, e.g. an applied
shear \cite{ElMasriJPCM2005}. Because in both cases the dynamics
may evolve on time scales comparable to or even shorter than the
relaxation time of $g_2$, a representation of the time evolution
of $c_I$ is more appropriate and insightful than that of the
two-time correlation function.
Limitations to the applicability of TRC are mainly due to the CCD
acquisition rate, which typically does not exceed a few tens or
hundreds of Hz. An additional experimental constraint is the need
to store the acquired images on the hard disk of a PC and to
process them at the end of the experiment: data sets up to several
Gb are not infrequent.

Other techniques have been proposed in the last few years to study
dynamical heterogeneity. New light scattering methods include
Speckle Visibility Spectroscopy (SVS)
\cite{DixonPRL2003,BandyopadhyayCondmat2005} and the measurement
of higher order intensity correlation functions
\cite{LemieuxJOSAA1999,LemieuxAPPOPT2001}. In a SVS experiment,
one measures $c_I(t,\tau=0)$, that is the instantaneous contrast
---or visibility--- of the speckle pattern,
$<I_{p}^2(t)>_p/<I_p(t)>_{p}^2 - 1$. The contrast depends on the
evolution of the speckle pattern during the exposure (integration)
time of the CCD. A significant evolution on this time scale leads
to a blurred speckle pattern image, and thus to a reduced
contrast. Fluctuations of the contrast can therefore be related to
dynamical fluctuations on the time scale of the exposure time.
Because the latter is typically much shorter than the time between
successive CCD acquisitions, SVS and TRC provide complementary
information, on fast and slow dynamics, respectively. Note that
SVS data can be obtained on the fly, with no need to store images,
since only one image at a time has to be processed. Higher order
intensity correlation functions
\cite{LemieuxJOSAA1999,LemieuxAPPOPT2001}, calculated by a
dedicated hardware, allow one to discriminate between homogeneous
and intermittent dynamics. The time scales that are probed and the
required measuring time are similar to those in a traditional
light scattering experiment: dynamics as fast as a fraction of
$\mu$sec can be measured, but the largest available delay is
limited to a few tens of seconds and the experiment duration has
to be at least 1000 times longer than the largest relaxation time
of the system. These constraints limit the applicability to glassy
soft matter.

Video and confocal scanning microscopy have been used to study
dynamical heterogeneity in concentrated colloidal suspensions
\cite{HabdasCOCIS2002}. Microscopy and TRC are complementary
techniques: while the former provides unsurpassed details on the
motion at a single particle level, scattering data typically
benefit from better statistics. Additionally, scattering
experiments usually require smaller particles than microscopy, a
plus when dealing with very slow dynamics and when sedimentation
effects should be minimized. Other techniques that can probe
heterogeneous behavior in soft glasses are dielectric
\cite{BuissonJPCM2003} and rheological \cite{BellonRSI2002}
measurements whose sensitivity is pushed to the limits set by
thermal fluctuations. The outcome of these experiments shares
intriguing similarities with TRC data, such as the non-Gaussian
distribution of voltage fluctuations in dielectric measurements on
suspensions of Laponite~\cite{BuissonJPCM2003}. Whether these
similarities are coincidental or stem from a common physical
origin remains an open question.

In the future, a deeper understanding of both spontaneous
dynamical fluctuations and the response to external perturbations
in soft glassy systems will likely require the combination of
different techniques, possibly applied simultaneously on the same
sample. We believe that TRC can play an important role in this
endeavor, thanks to its ability to detect instantaneous variations
in the dynamics.

\acknowledgements We thank P. Ballesta, E. Pitard, L. Berthier, P.
Holdsworth, J. P. Garrahan  for many useful discussions, and D.
Popovic for bringing to our attention the works on the second
spectrum. This work is supported in part by the European MCRTN
``Arrested matter'' (MRTN-CT-2003-504712) and the NoE
``Softcomp'', and by CNES (grant no. 03/CNES/4800000123), CNRS
(PICS no. 2410), Minst\`{e}re de la Recherche (ACI JC2076) and the
Swiss National Science Foundation. L.C. is a junior member of the
Institut Universitaire de France, whose support is gratefully
acknowledged.

\appendix
\section{Calculation of the variance and covariance terms in Equation~(\ref{Eq:sigmagen})}

In this section we provide additional details on the two
approaches to the calculation of the variance of the measurement
noise mentioned in sec. \ref{SEC:fluctnoise}. We first show that
the variance and covariance terms in Eq.~(\ref{Eq:sigmagen}) may
be expressed in a way that is independent of the homogeneous $vs.$
heterogeneous nature of the dynamics, and then explicitly
demonstrate the $N^{-1}$ scaling of $\sigma^2_{n}$ that was used
in the correction scheme presented in this paper.

The quantities of interest are the variance of the pixel-averaged
intensity, $\sigma^{2}_{I}$, the covariance between $I$ and $J$,
$\sigma_{I,J}(\tau)$, the variance of the un-normalized intensity
correlation function, $\sigma^{2}_{G_2}(\tau)$, and  the
covariance between $G_2$ and $I$, $\sigma_{G_2,I}(\tau)$. We
recall that $I(t) = <I_p(t)>_p$, $J(t) = <I_p(t+\tau)>_p$ and drop
in the following the explicit dependence on $t$ in the notation of
all time-varying variables. We assume the dynamics to be
stationary and the experiment duration much longer than the
average relaxation time of $c_I$. We start by noting that
$\sigma^{2}_{I}$ quantifies the fluctuations of the
$instantaneous$ pixel-averaged intensity. Therefore,
$\sigma^{2}_{I}$ is independent of the nature of the dynamics:
indeed, any dynamical process that fully renews the speckle
pattern will allow all possible values of $I$ to be sampled over
time and thus will lead to the same $\sigma^{2}_{I}$. In contrast,
$\sigma_{I,J}(\tau)$, $\sigma^{2}_{G_2}(\tau)$, and
$\sigma_{G_2,I}(\tau)$ depend on the way the correlation between
$distinct$ speckle patterns fluctuates with time and therefore
contain contributions due both to the measurement noise and to the
intrinsic fluctuations of the dynamics. However, note that the
$\tau \rightarrow 0$ and $\tau \rightarrow \infty$ limits of these
quantities are insensitive to the dynamics, because they involve
either instantaneous properties of the speckle pattern (for
$\tau=0$), or quantities related to pairs of speckle images
totally uncorrelated (for $\tau \rightarrow \infty$). This
observation, together with the linear dependence of
$\sigma_{I,J}(\tau)$, $\sigma^{2}_{G_2}(\tau)$, and
$\sigma_{G_2,I}(\tau)$ on $\overline{c_I}$ in the absence of
dynamical heterogeneity shown in fig. \ref{Fig:sigma}, suggests
the following form for the contribution of the the measurement
noise to the covariance and variance terms:
\begin{eqnarray}
\sigma_{I,J}(\tau)  =  \sigma_{I,J}(\infty) +
\frac{\overline{c_I(\tau)}}{\beta} \left[\sigma_{I,J}(0) -
\sigma_{I,J}(\infty)\right] \label{Eq:AppIa} \nonumber \\
 \sigma^2_{G_2}(\tau)  =   \sigma^2_{G_2}(\infty) +
\frac{\overline{c_I(\tau)}}{\beta} \left[\sigma^2_{G_2}(0) -
\sigma^2_{G_2}(\infty)\right] \label{Eq:AppIb} \nonumber
\\
  \sigma_{G_2,I}(\tau)   =  \sigma_{G_2,I}(\infty)  \nonumber
\\
 + \frac{\overline{c_I(\tau)}}{\beta} \left[\sigma_{G_2,I}(0) -
\sigma_{G_2,I}(\infty)\right]. \label{Eq:AppIc}
\end{eqnarray}
All coefficients in the r.h.s. of the above equations are not
affected by dynamical heterogeneity. Table I summarizes their
values in terms of quantities that can be directly obtained from
the speckle images, regardless of the nature of the dynamics.

\begin{table}
\caption{\label{tab:table1} Coefficients needed to calculate the
contribution of the measurement noise to $\sigma_{I,J}(\tau)$,
$\sigma^{2}_{G_2}(\tau)$, and $\sigma_{G_2,I}(\tau)$ .}
\begin{ruledtabular}
\begin{tabular}{ccc}
&$\tau = 0$&$\tau \rightarrow \infty$\\
\hline
$\sigma_{I,J}(\tau)$ & $\sigma^2_{I}$ & 0\\
$\sigma^2_{G_2}(\tau)$ & $ \sigma^2_{G_2(0)}$& $2\overline{I}^2\sigma^2_{I}$\\
$\sigma_{G_2,I}(\tau)$ & $\sigma_{G_2(0),I}$ &$\overline{I}^2\sigma^2_{I}$ \\

\end{tabular}
\end{ruledtabular}
\end{table}

As mentioned in sec.~\ref{sec:noisedeconvolution}, estimates of
$\sigma^2_{n}$ obtained by using Eq.~(\ref{Eq:AppIc}) and Table I
are typically affected by a significant error. In this paper we
have proposed a more robust correction method, based on the
$N^{-1}$ scaling of the measurement noise variance, which we
demonstrate here. For the sake of simplicity, we assume in the
following that there is no correlation between the instantaneous
value of the intensity at distinct pixels, i.e. $\overline{I_p
I_q} = \overline{I_p} ~\overline{I_q}$ for $p \neq q$. Physically,
this corresponds to a speckle size much smaller than the pixel
size; this requirement considerably simplifies the calculations,
but it can be relaxed without changing the $N^{-1}$ scaling, as we
shall show at the end of this Appendix.

For experiments whose duration is much longer than the relaxation
time of $\overline{c_I}$, the intensity at any given pixel fully
fluctuates many times and its probability distribution over time
is the same as the instantaneous PDF of $I_p$ calculated over all
$N$ pixels; therefore averages over time and over pixels can be
swapped. We take advantage of this property and of the statistical
independence between $I_p$ and $I_q$ to write
\begin{eqnarray}
 \sigma^2_{I}  =  \overline{\frac{1}{N^2}\sum_{p,q}{I_pI_q}} -
 \overline{I}^2\nonumber\\
 = \frac{1}{N^2}\sum_{p=1}^{N}{\overline{I_{p}^2}} + \frac{1}{N^2}\sum_{p\neq q}{\overline{I_p}\,\overline{I_q}}-
 \overline{I}^2.
\label{Eq:Appsigma2I}
\end{eqnarray}
This expression may be further simplified by noting that moments
of the intensity at any pixel are equal to those of, let's say,
pixel 1: $\overline{I_{p}^2} = \overline{I_{1}^2}$ and
$\overline{I_{p}} = \overline{I_1} = \overline{I}$. Hence
\begin{eqnarray}
 \sigma^2_{I}  = \frac{1}{N}\overline{I_{1}^2} + \frac{N(N-1)}{N^2}\overline{I}^2 - \overline{I}^2 =
 \frac{1}{N}\overline{I_{1}^2}.
\label{Eq:Appsigma2I2}
\end{eqnarray}
Similarly, one obtains
\begin{eqnarray}
 \sigma^2_{I,J}  =  \overline{\frac{1}{N^2}\sum_{p,q}{I_pJ_q}} -
 \overline{I}\,\overline{J}\nonumber\\
 = \frac{1}{N}\overline{G_2} + \frac{N(N-1)}{N^2}\overline{I}\,\overline{J} - \overline{I}\,\overline{J} = \frac{1}{N}\overline{c_I}
 \overline{I}^2,
\label{Eq:AppsigmaIJ}
\end{eqnarray}
\begin{eqnarray}
 \sigma_{G_2,I}  =  \overline{\frac{1}{N^2}\sum_{p,q}{I_pJ_pI_q}} -
 \overline{G_2}\,\overline{I}\nonumber\\
 = \frac{1}{N}\overline{I_{1}^2J_1} - \frac{1}{N}\overline{G_2}
 \overline{I},
\label{Eq:AppsigmaG2I}
\end{eqnarray}
and
\begin{eqnarray}
 \sigma^2_{G_2}  =  \overline{\frac{1}{N^2}\sum_{p,q}{I_pJ_pI_qJ_q}} -
 \overline{G_2}^2\,\nonumber\\
 = \frac{1}{N}\overline{I_{1}^2J_{1}^2} -
 \frac{1}{N}\overline{G_2}^2.
\label{Eq:Appsigma2G2}
\end{eqnarray}
Equations (\ref{Eq:Appsigma2I2})-(\ref{Eq:Appsigma2G2}) show that
all the terms in the l.h.s. of the expression of $\sigma_{n}^2$,
Eq.~(\ref{Eq:sigmagen}), are indeed proportional to $N^{-1}$.

As a final remark, we show that the form of the equations derived
above is not changed by a short-ranged correlation between the
intensity of distinct pixels, such as that typically observed in
the CCD speckle images. We consider, as an example, the
calculation of $\sigma^{2}_{I}$; similar arguments apply also to
the other variance and covariance terms. Because the spatial
correlation is short-ranged, the intensity of pixel $p$ will be
correlated only to that of a small number, $M$, of nearby pixels,
whereas for all other pixels $\overline{I_pI_q} \approx
\overline{I_p}\,\overline{I_q}$. We indicate the set of nearby
pixels by $Q_p$ and split the double sum over distinct pixels in
Eq.~(\ref{Eq:Appsigma2I}) as follows:
\begin{eqnarray}
 \frac{1}{N^2}\sum_{p\neq q}{\overline{I_pI_q}} =
 \frac{1}{N^2}\sum_{p,
 q \in Q_p}{\overline{I_pI_q}}+\frac{1}{N^2}\sum_{p,q \not\in
 Q_p}{\overline{I_p}\,\overline{I_q}}
 \nonumber\\
 \approx \frac{1}{N}\sum_{ q \in Q_1}{\overline{I_1I_q}} +
 \frac{N(N-M)}{N^2}\overline{I}.
\label{Eq:Apppneqq}
\end{eqnarray}
By substituting Eq.~(\ref{Eq:Apppneqq}) in
Eq.~(\ref{Eq:Appsigma2I}), one finds that $\sigma^{2}_{I}$ still
scales as $N^{-1}$, although with a different prefactor:
\begin{eqnarray}
\sigma^{2}_{I} = \frac{1}{N}\overline{I_{1}^2} +
 \frac{1}{N}\sum_{ q \in Q_1}{\overline{I_1I_q}} -
 \frac{M}{N}\overline{I}^2.
 \label{Eq:Appsigma2I3}
\end{eqnarray}

\section{PDF of $c_I$}

We show here that in the limit of large $N$ and for homogeneous
dynamics the PDF of $c_I$ is Gaussian, as found experimentally
(see fig.~\ref{Fig:PDFBrownian}). Additionally, we will
demonstrate the relationship $\overline{G_2} =
\overline{I}^2(\overline{c_I}+1)$. Note that this relationship is
not trivial, since in general $(\overline{c_I}+1) =
\overline{\left[\frac{<I_pJ_p>_p}{IJ} \right]}$ differs from
$\frac{\overline{<I_pJ_p>_p}}{\overline{I}\,\overline{J}} =
\frac{\overline{G_2}}{\overline{I}^2}$.

We start by noting that $G_2$, $I$, and $J$ are obtained from an
average over a large number $N$ of pixels. Because of the central
limit theorem, their PDF is Gaussian, with a standard deviation
much smaller than the mean. Accordingly, at any time $G_2(t) =
\overline{G_2}\left [ 1 + \epsilon_{G_2}(t)\right]$, with
$\epsilon_{G_2}$ a Gaussian distributed random variable with mean
$\overline{\epsilon_{G_2}} = 0$ and variance  $\sim N^{-1}$. One
can write similar expressions for $I(t)$ and $J(t)$, so that to
leading order
\begin{eqnarray}
   \frac{G_2}{IJ}  =  \frac{\overline{G_2}(1+\epsilon_{G_2})}{\overline{I}(1+\epsilon_{I})\overline{J}(1+\epsilon_{J})}
\approx   \frac{\overline{G_2}} {\overline{I}^2} (1+\xi),
   \label{Eq:AppG2overIJ2}
\end{eqnarray}
where we have used $\overline{J}=\overline{I}$ and have introduced
$\xi = \epsilon_{G_2}-\epsilon_{I}-\epsilon_{J}$. $\xi$ is the sum
of three (partially correlated) Gaussian random variables and
therefore is itself Gaussian distributed \cite{Frieden2001}, with
mean
$\overline{\epsilon_{G_2}}-\overline{\epsilon_{I}}-\overline{\epsilon_{J}}
= 0$. It follows that the PDF of
\begin{eqnarray}
   c_I  =  \frac{G_2}{IJ}-1 \approx \frac{\overline{G_2}} {\overline{I}^2} (1+\xi) - 1
   \label{Eq:AppPDFcI}
\end{eqnarray}
is Gaussian, with mean $\overline{c_I} = \frac{\overline{G_2}}
{\overline{I}^2}- 1$. Thus, $\overline{G_2} =
\overline{I}^2(\overline{c_I}+1)$.


\end{document}